\documentclass[12pt,preprint]{aastex61}

\newcommand{\feka}{\hbox{Fe\,K$\alpha$}}

\newcommand{\lumin}{\hbox{erg~s$^{-1}$}}

\newcommand{\be}{\begin{equation}}
\newcommand{\ee}{\end{equation}}
\newcommand{\ba}{\begin{eqnarray}}
\newcommand{\ea}{\end{eqnarray}}

\newcommand{\chandra}{{\emph{Chandra}}}
\newcommand{\cxo}{{\emph{Chandra X-ray Observatory}}}
\newcommand{\cxc}{{\emph{Chandra X-ray Center}}}

\newcommand{\nasa}{\emph{NASA}}
\newcommand{\xspec}{\emph{XSPEC}}
\newcommand{\QJ}{Q\,J0158$-$4325}
\newcommand{\HEone}{HE\,0435$-$1223}
\newcommand{\SD}{SDSS\,J1004$+$4112}
\newcommand{\HEtwo}{HE\,1104$-$1805}
\newcommand{\Q}{Q\,2237$+$0305}

\newcommand{\simgt}{\lower 2pt \hbox{$\, \buildrel {\scriptstyle >}\over {\scriptstyle\sim}\,$}}
\newcommand{\simlt}{\lower 2pt \hbox{$\, \buildrel {\scriptstyle <}\over {\scriptstyle\sim}\,$}}
\newcommand{\ls}{\lower 2pt \hbox{$\;\scriptscriptstyle \buildrel<\over\sim\;$}}
\newcommand{\gs}{\lower 2pt \hbox{$\;\scriptscriptstyle \buildrel>\over\sim\;$}}

\newcommand{\sar}{$^{\prime\prime}\!\!.$}

\begin{document}

\title{Constraining Quasar Relativistic Reflection Regions and Spins with Microlensing}
\author{Xinyu Dai}
\affiliation{Homer L. Dodge Department of Physics and Astronomy,
University of Oklahoma, Norman, OK, 73019}
\email{xdai@ou.edu, ss507@ou.edu}
\author{Shaun Steele}
\affiliation{Homer L. Dodge Department of Physics and Astronomy,
University of Oklahoma, Norman, OK, 73019}
\author{Eduardo Guerras}
\affiliation{Homer L. Dodge Department of Physics and Astronomy,
University of Oklahoma, Norman, OK, 73019}
\author{Christopher W. Morgan}
\affiliation{Department of Physics, United States Naval Academy, 572C Holloway Road, Annapolis, MD 21402, USA}
\author{Bin Chen}
\affiliation{Department of Scientific Computing, Florida State University, Tallahassee, FL 32306,
USA}

\begin{abstract}
We present an analysis of \chandra\ spectra of five gravitationally lensed active galactic nuclei. We confirm the previous detections of FeK$\alpha$ emission lines in most images of these objects with high significance. The line energies range from 5.8 to 6.8 keV with widths from unresolved to 0.6 keV, consistent with emission close to spinning black holes viewed at different inclination angles.
We also confirm the positive offset from the Iwasawa-Taniguchi effect, the inverse correlation between the FeK$\alpha$ equivalent width and the X-ray luminosity in AGN, where our measured equivalent widths are larger in lensed quasars.  We attribute this effect to microlensing, and perform a microlensing likelihood analysis to constrain the emission size of the relativistic reflection region and the spin of supermassive black holes, assuming that the X-ray corona and the reflection region, responsible for the iron emission line, both follow power-law emissivity profiles.
The microlensing analysis yields strong constraints on the spin and emissivity index of the reflection component for \Q, with $a > 0.92$ and $n > 5.4$.  For the remaining four targets, we jointly constrain the two parameters, yielding $a=0.8\pm0.16$ and   
an emissivity index of $n=4.0\pm 0.8$, suggesting that the relativistic X-ray reflection region is ultra-compact and very close to the innermost stable circular orbits of black holes, which are spinning at close to the maximal value.  We successfully constrain the half light radius of the emission region to  $< 2.4$ $r_g$ ($r_g = GM/c^2$) for \Q\ and in the range 5.9--7.4 $r_g$ for the joint sample.
\end{abstract}


\section{Introduction}
The X-ray spectra of active galactic nuclei (AGN) are characterized by continuum emission that is well modeled by a power law \citep[e.g.,][]{guilbert1988, reynolds2003, brenneman2013}. The UV emission of the accretion disk provides the seed photons and these photons are then inverse Compton scattered by relativistic electrons in the corona to produce the continuum. A portion of these photons are scattered back to the accretion disk and can can create a reprocessed or reflected emission component including fluorescent emission lines, most notably, the \feka\ emission line at 6.4\,keV in the source rest frame \citep[]{guilbert1988, fabian1995, reynolds2003, brenneman2013}. 
The exact locations of the reflection are not well constrained and the process can occur at multiple locations, from the inner accretion disk, disk, broad line regions, to the torus. 

Measuring the spins of super-massive black holes (SMBH) at the centers of AGN is important because it is related to the growth history of the black holes, their interaction with the environment, the launching of relativistic jets, and the size of the innermost stable circular orbit (ISCO) \citep[e.g.,][]{thorne1974, blandford1977, fabian2012, brenneman2013, chartas2017}. For example, as a SMBH grows it can provide matter and energy to its surrounding environment through outflows \citep{fabian2012}. One important method to estimate spins models the general relativistic (GR) and special relativistic (SR) distortions of the \feka\ emission line \citep[e.g.,][]{brenneman2013, reynolds2014}. This method has been applied to many nearby Seyferts with most estimates being close to the maximal spin ~\citep{reynolds2014}. Another approach is to model the UV-optical SEDs of high redshift quasars \citep[e.g.,][]{capellupo2015, capellupo2017}. These studies again find high spins for high redshift quasars. 

Quasar microlensing has significantly improved our understanding of the accretion disks \citep[e.g.,][]{dai2010, morgan2010, mosquera2013, blackburne2014, blackburne2015, macleod2015} and non-thermal emission regions \citep[e.g.,][]{pooley2006, pooley2007, morgan2008, chartas2009, chartas2016, chartas2017, dai2010, chen2011, chen2012, guer2017} of quasars, and the demographics of microlenses in the lens galaxy \citep[e.g.,][]{blackburne2014, dg2018, gdm2018} . Since the magnification diverges on the caustics produced by the lensing stars, quasar microlensing can constrain arbitrarily small emission regions if they can be isolated from other emission, in position, velocity, or energy. In particular, microlensing can be used to constrain the spin of black holes by measuring the ISCO size.

In this paper, we will utilize the excess equivalent width (EW) difference between lensed and unlensed quasars first summarized by \citet{chen2012} to constrain the size of the reflection region and the spin of quasars. This paper is organized as follows.  We present the \chandra\ observations and the data reduction in Section~\ref{sec:obs_red} and the spectral analysis in Section~\ref{sec:spec}. In Section~\ref{sec:spec} we discuss the significance of the iron line detections and we confirm the offset of \feka\ equivalent widths of lensed quasars. In Section~\ref{sec:ML}, we carry out a microlensing analysis to estimate the size of the \feka\ emission region and the spin of the black hole. We discuss the results in Section~\ref{sec:discussion}.
We assume cosmological parameters of $\Omega_M = 0.27$, $\Omega_{\Lambda}=0.73$, and $H_0 = 70$~km~s$^{-1}$~Mpc$^{-1}$ throughout the paper.

\section{OBSERVATIONS AND DATA REDUCTION}
\label{sec:obs_red}
The observations used in this paper were made with the  Advanced CCD Imaging Spectrometer \citep{garmire2003} on board \cxo\ \citep{weisskopf2002}. \chandra\ has a point spread function (PSF) of 0\sar5 and is therefore able to resolve most of the multiple images in lensed systems since they have a typical image separation of 1--2\arcsec. We used two sets of data in this analysis. The first, which we call Data Set 1, mainly comes from the \chandra\ Cycle 11 program and the second, which we call Data Set 2, comes from \chandra\ Cycles 14--15. We analyzed five lenses: \QJ, \HEone, \SD, \HEtwo, and \Q. The lens properties are summarized in Table~\ref{tab:lensinfo}. 
All the data were reprocessed using the \cxc\ CIAO 4.7 software tool \texttt{chandra\_repro}, which takes data that have already passed through the \cxc\ Standard Data Processing and filters the event file on the good time intervals, grades, cosmic ray rejection, transforms to celestial coordinates, and removes any observation-specific bad pixel files. 

\begin{deluxetable}{lccccccccc}
	\tabletypesize{\scriptsize}
    \caption{Gravitational Lenses Analyzed in This Paper \label{tab:lensinfo}}
    \tablewidth{0pt}
    \tablehead{
      \colhead{Object} & \colhead{\(z_s\)} & \colhead{\(z_l\)} & \colhead{R.A.(J2000)} & \colhead{Dec.(J2000)} & \colhead{Galactic \(N_H\)} & \colhead{Epochs} & \multicolumn{3}{c}{\uline{\chandra\ Expo.\ Time}} \\
    \colhead{} & \colhead{} & \colhead{} & \colhead{} & \colhead{} & \colhead{} & \colhead{} & \colhead{Data Set I} & \colhead{Data Set II} & \colhead{Combined} \\
      \colhead{} & \colhead{} & \colhead{} & \colhead{} & \colhead{} & \colhead{($10^{20}$ cm$^{-2}$)} & \colhead{} & \multicolumn{3}{c}{($10^3$~sec)} 
    }
    \startdata
    QJ 0158$-$4325 & 1.29 & 0.317 & 01:58:41.44 & $-$43:25:04.20 & 1.95 & 12   & 29.9  & 111.7 & 141.6 \\ 
    HE 0435$-$1223 & 1.689 & 0.46 & 04:38:14.9 & $-$12:17:14.4 & 5.11 & 10     &  48.4 & 217.5 & 265.9 \\ 
    SDSS J1004$+$4112 & 1.734 & 0.68 & 10:04:34.91 & $+$41:12:42.8 & 1.11 & 11 & 103.8 & 145.6 & 249.4 \\  
    HE 1104$-$1805 & 2.32 & 0.73 & 11:06:33.45 & $-$18:21:24.2 & 4.62 & 15     & 110.0 &  80.5 & 191.5 \\  
    Q 2237$+$0305 & 1.69 & 0.04 & 22:40:30.34 & $+$03:21:28.8 & 5.51 & 30      & 292.4 & 175.4 & 467.8 \\
    \enddata
\end{deluxetable}    
\section{SPECTRAL ANALYSIS}
\label{sec:spec}
We extracted spectra using the CIAO 4.7 software tool \texttt{specextract}. We used circular extraction regions with a radius of 0\sar72, which is less than the typical image separation but greater than the PSF of \chandra. We chose this extraction radius to balance the needs for maximizing the  signal-to-noise ratio (S/N) of the spectrum from a lensed image while minimizing the contamination from the other nearby lensed images. For the cluster lens, SDSS J1004+4112, we instead used circles with a radius 1\sar5, because the lensed images are well separated in this system and are therefore not contaminated by the other images. The extraction regions were centered on the positions found from the PSF fits performed in \citet{guer2017} to the X-ray images. For the background regions, a circular region with radius of 0\sar72 was reflected through the position of the other images of the lens to account for both large scale backgrounds and any contamination from the other images. For SDSS J1004+4112, we used partial annuli to account for the non-negligible X-ray contamination from the X-ray emission of the cluster that acts as the lens for this object. We also extracted ``Total'' spectra using circular regions that encompass all images. Again, SDSS J1004+4112 was treated differently and the ``Total" spectra is the sum of the individual image spectra. These regions are similar to those used in \citet{chen2012}, where a spectral analysis was performed for Data Set 1. \citet{chartas2017} also performed an analysis on individual epochs of RX\,J1131$-$1231, \QJ, and \SD.

\begin{figure}
\centering
\epsscale{0.7}
\plotone{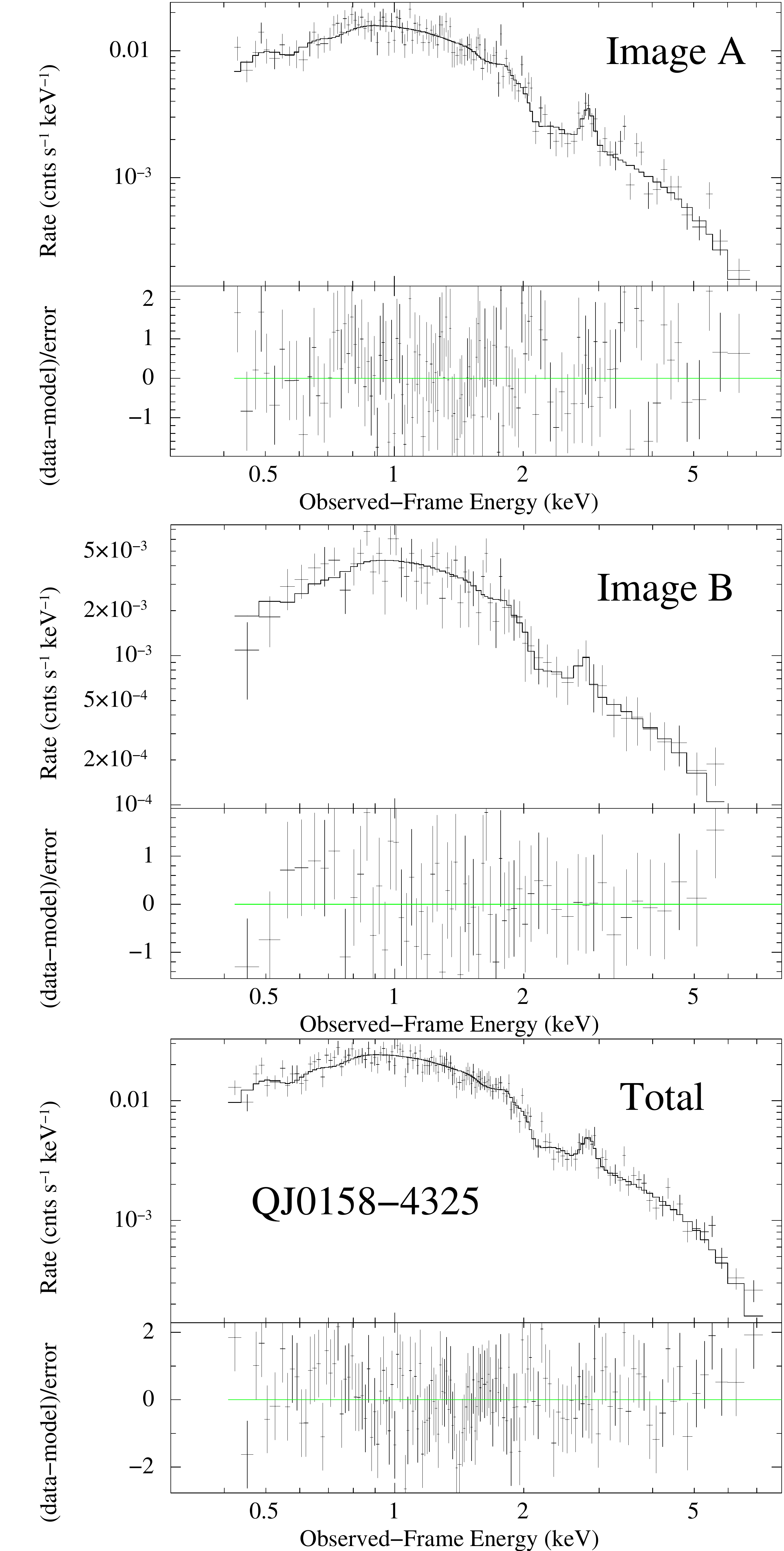}
\caption{Combined spectra of \QJ\ along with the best power-law plus gaussian emission line fit and its residuals. Images A and Total have  significant ($>99\%$) line detections.\label{fig:0158_spec}}
\end{figure}

\begin{figure}
\centering
\epsscale{1.2}
\plotone{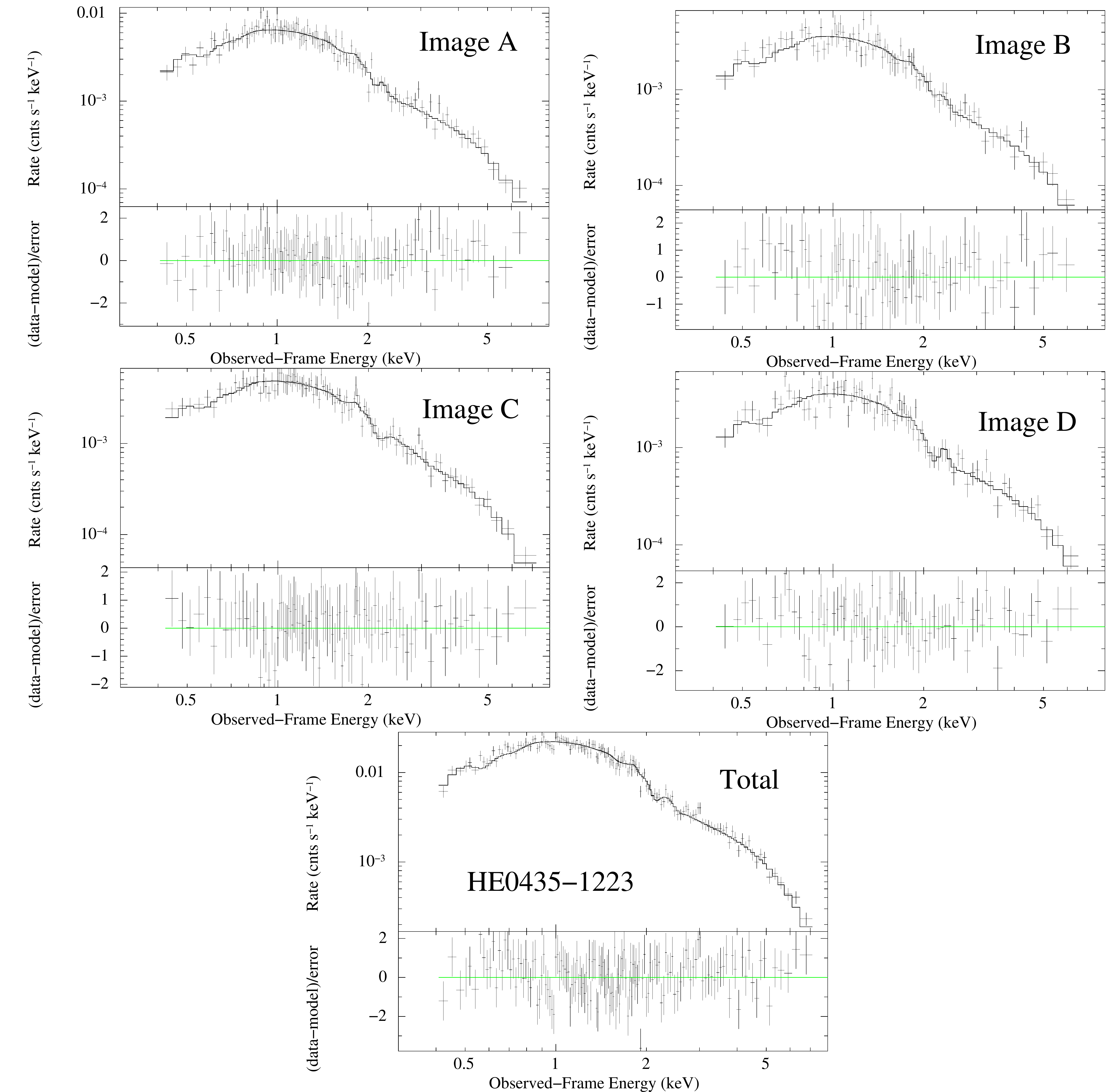}
\caption{Combined spectra of \HEone\ along with the best power-law plus gaussian emission line fit and its residuals. Image C and Total have  significant ($>99\%$) line detections.\label{fig:0435_spec}}
\end{figure}

\begin{figure}
\centering
\epsscale{1.2}
\plotone{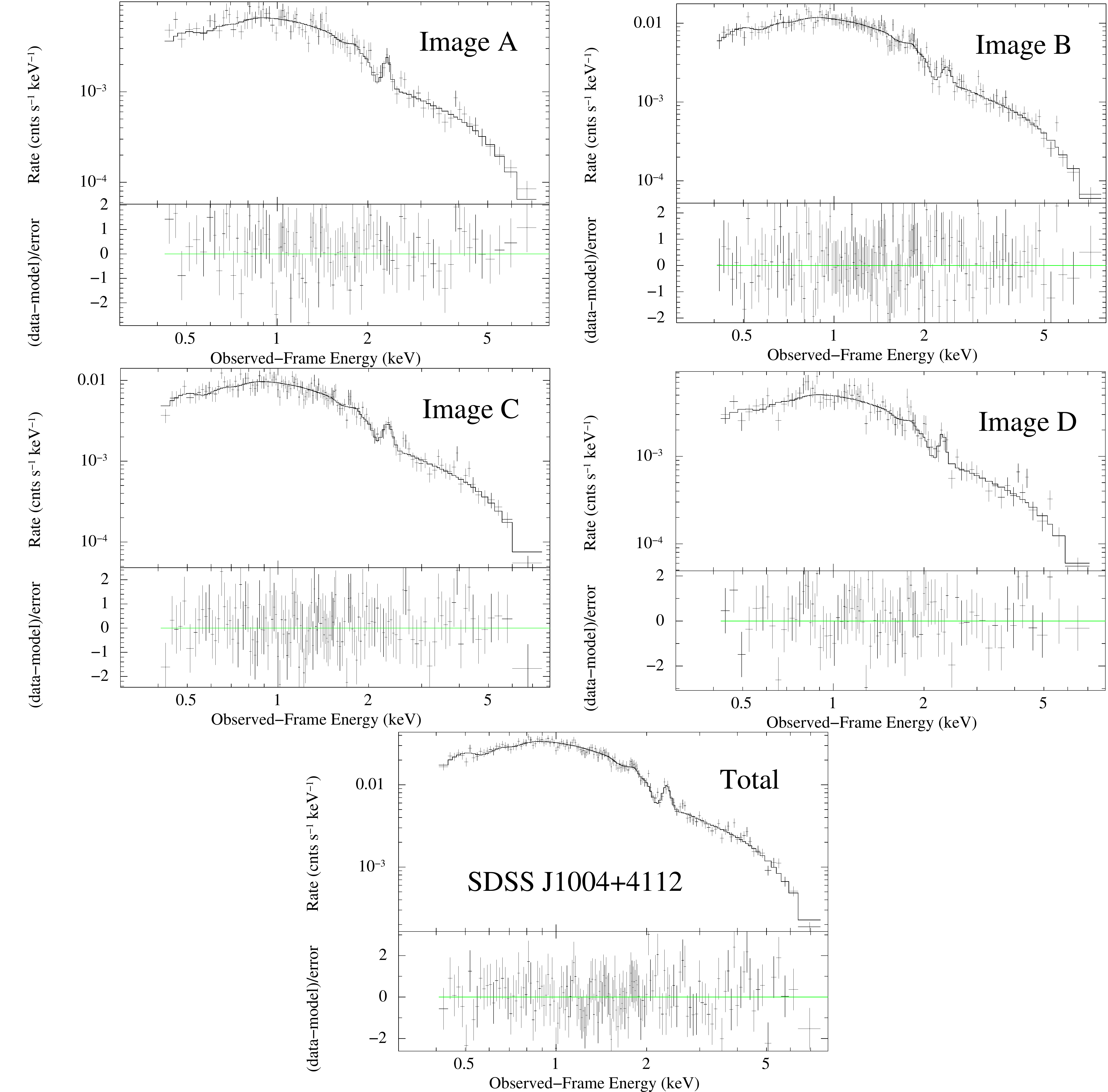}
\caption{Combined spectra of \SD\ along with the best power-law plus gaussian emission line fit and its residuals. All images of \SD\ have  significant ($>99\%$) line detections.\label{fig:1004_spec}}
\end{figure}

\begin{figure}
\centering
\epsscale{0.7}
\plotone{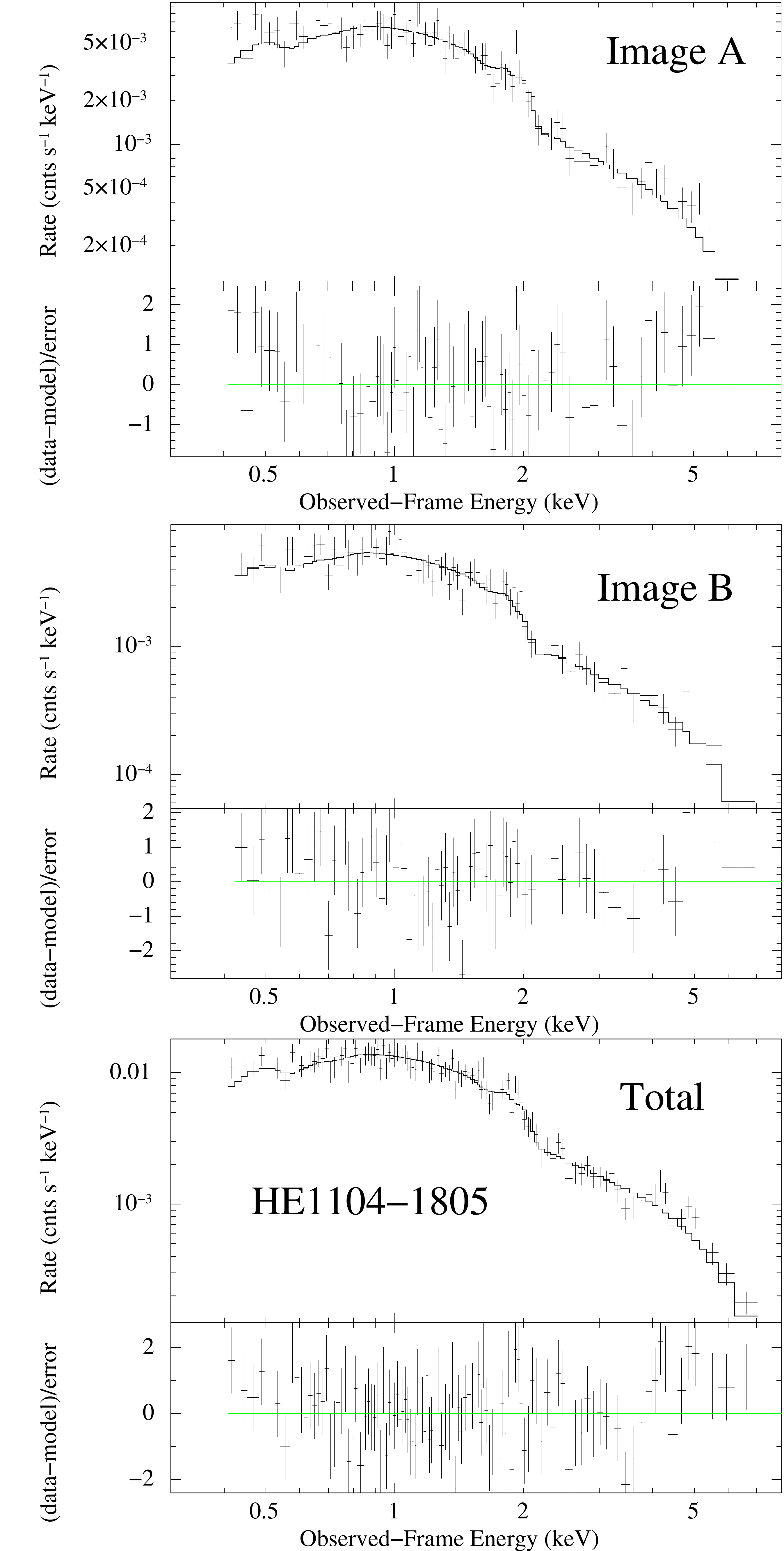}
\caption{Combined spectra of \HEtwo\ along with the best power-law plus gaussian emission line fit and its residuals. None of the images have a significant ($>95\%$) line detection.\label{fig:1104_spec}}
\end{figure}

\begin{figure}
\centering
\epsscale{1.2}
\plotone{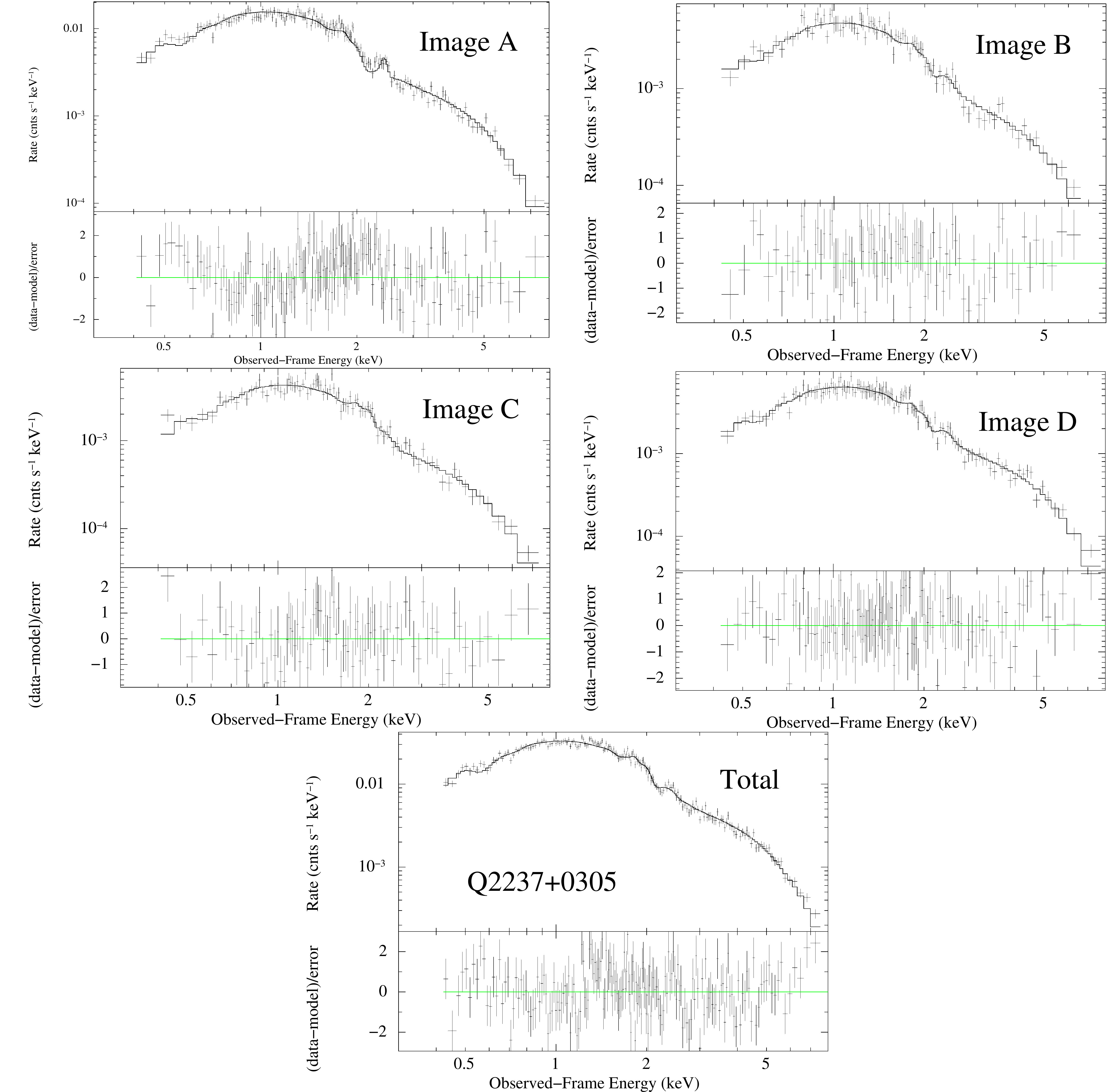}
\caption{Combined spectra of \Q\ along with the best power-law plus gaussian emission line fit and its residuals. All images of \Q\ have  significant ($>99\%$) line detections.\label{fig:2237_spec}}
\end{figure}

We fit the extracted spectra using the \nasa\ \emph{HEASARC} software \xspec \emph{V12.9}. We used a simple power-law model for the direct X-ray continuum and then added Gaussian components for any emission lines. These were modified by Galactic absorption \citep{dickey90} and absorption from the lens. Galactic absorption was fixed for each system to the values given in Table~\ref{tab:lensinfo}. The lens absorption was a free parameter in the fitting process, unless there was no evidence for absorption in the lens. In these cases, the lens absorption was set to zero. The source and lens redshifts are also listed in Table~\ref{tab:lensinfo}. The spectral fitting results for combined data, Data Set 1, and Data Set 2 are given in Tables~\ref{tab:spec_fit_results},~\ref{tab:spec_fit_results_1}, and~\ref{tab:spec_fit_results_2} respectively. In these tables, we report the photon index \(\Gamma\), the lens absorption \(N_H\), the Gaussian line properties (rest frame line energy, width, and EW)\footnote{The rest frame EW is calculated using the XSPEC command eqw and then times $(1+z)$ to correct the cosmological redshift effect.}, the reduced $\chi^2$ of the fit, the null-hypothesis probability of the fit, the analytical and Monte Carlo calibrated significances of the metal line, the line detection threshold of the spectrum, and the absorption free and macro magnification corrected X-ray luminosity \(L_{X}\) over the 10--50\,keV rest frame range. We fit the spectra in the 0.4--8.0\,keV observed frame. The \xspec\ command \emph{dummyrsp} was used to extend our model to 50\,keV. The magnifications were calculated using the equation 
\begin{equation}\label{eqn:mag_eqn}
\mu= |(1-\kappa)^2-\gamma^2|^{-1},
\end{equation}
where \(\kappa\) and \(\gamma\) are the surface mass density and shear for each image, and we adopted the values from Table~9 of \citet{guer2017}.  The reported luminosity values are different for individual images of each lens, because of the combination of source variability, time-delay, and microlensing effects. The spectral fits and \(\Delta\chi ^2\) plots are  shown in Figures~\ref{fig:0158_spec}--\ref{fig:2237_spec}.

We detected the FeK\(\alpha\) emission line at high significances ($>99\%$) in the combined, ``Total'' spectra in four out of five lenses and with a weaker detection in \HEtwo\ (see Table~\ref{tab:spec_fit_results}). For the combined data sets of the individual images, we detected the FeK\(\alpha\) line with high significance ($>99\%$) in ten images of the combined data sets, and we detected low significance FeK\(\alpha\) lines in \QJ\ image B, \HEone\ images A, B, and D, and \HEtwo\ image A (see Table~\ref{tab:spec_fit_results}). We were unable to obtain a stable fit for \HEtwo\ image B that included the iron emission line.

Analyzing the data sets separately, we detected the FeK\(\alpha\) emission line in the ``Total" image for \SD\ and \Q\ with $>99\%$ significance for Data Set 1. We have a weak detection in the ``Total" image for \QJ, \HEone, and \HEtwo\ (see Table~\ref{tab:spec_fit_results_1}). For the individual images, we find $>99\%$ significant detections only in the images of \SD\ and \Q. The rest have weak detections except for \QJ\ image B, \HEone\ image D, and \HEtwo\ image B where we were unable to obtain stable fits for that included the iron emission line.  Compared to the analysis of \citet{chen2012} of Data Set 1, the line detections and significance values are generally consistent, however, with the significance values reported in this paper slightly more conservative.

We detected the FeK\(\alpha\) emission line with $>99\%$ significance in the ``Total" image in four of the five lenses for Data Set 2, except \HEtwo\ with a significance value of 96\% (see Table~\ref{tab:spec_fit_results_2}). We report weak detections in \QJ\ image B, \HEone\ images B and D, and \Q\ image D. For \HEtwo\ image B, we were unable to obtain a stable fit that included the iron emission line. The non-detection of the iron line in Data Set 1 for \QJ\ image B or \HEone\ image D is in agreement with \citet{chen2012}, and we have weak line detections in these two images in Data Set 2 and in the combined data set as well. 

\begin{figure}
\epsscale{1.}
\plotone{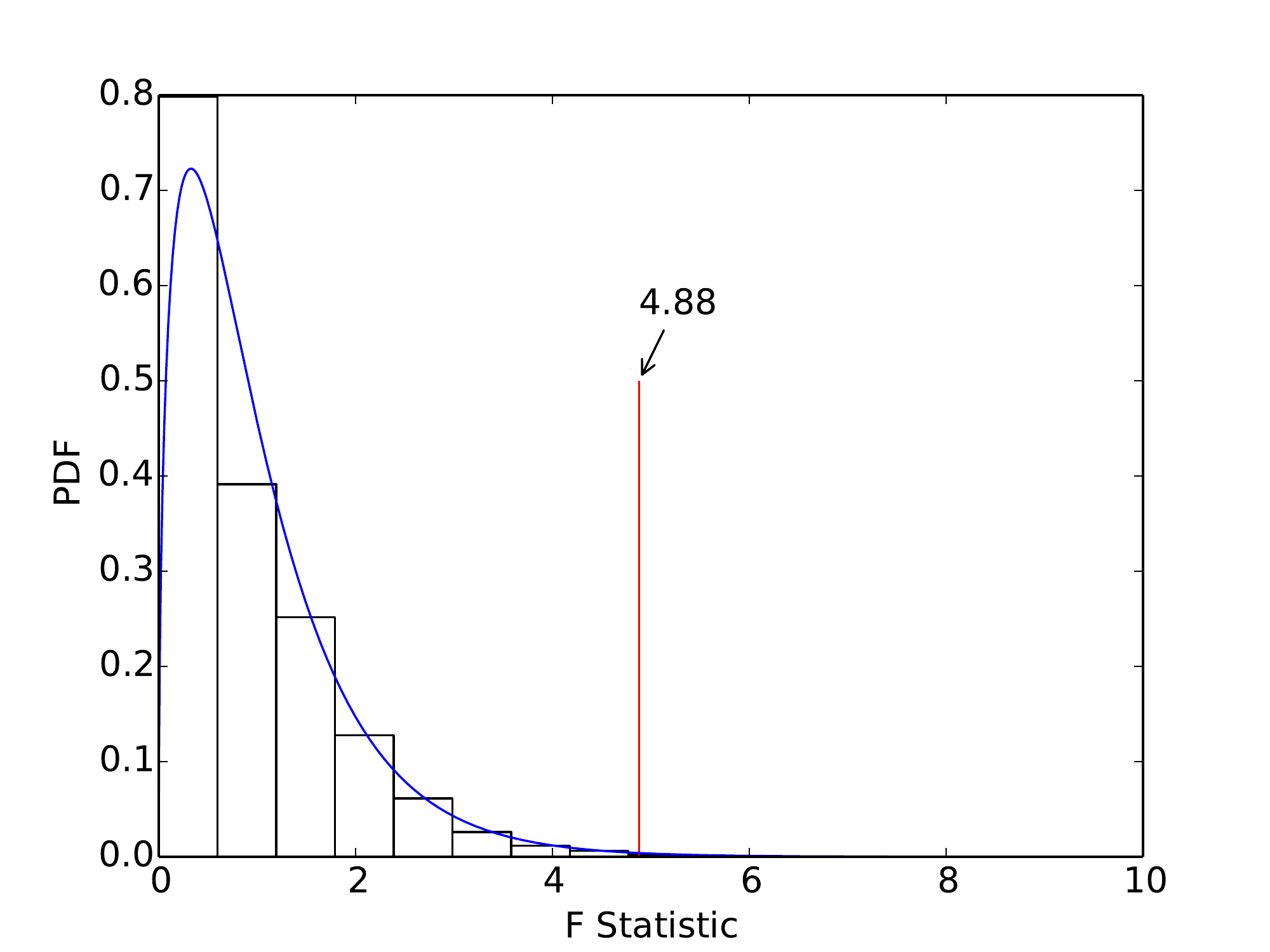}\caption{Significance of the \feka\ line detection for \QJ\ image A. The black histogram shows the null probability distribution of F-statistics from Monte Carlo simulations, the analytical distribution is the blue curve, and the F statistic from the real data is shown by the red vertical line. We see here that the Monte Carlo results follow the analytical curve closely for this example, and the measured F-statistic (red vertical line) is larger than any of the Monte Carlo values giving it a 99.99\% significance.
}
\label{fig:fdist}
\end{figure}

\citet{protassov2002} showed that the F-test should be used with caution when testing the significance of  emission line detections since the data may not necessarily follow an F-distribution. If the models are well constrained, \citet{protassov2002} provided a method to calibrate the F statistic using Monte Carlo simulations of the spectra. Following this method and the example of \citet{dai2003}, we generated ten thousand simulated spectra for each image of each object in all data sets to evaluate the significance of the detected emission lines. We used the \xspec\ command \emph{fakeit} to simulate the spectra. We fixed the null model parameters (not including the emission line) to the best fit values listed in Tables~\ref{tab:spec_fit_results}--\ref{tab:spec_fit_results_2}. These simulated spectra were then grouped exactly like the real data and fit with the null model and the model with the line. 
From the fits, we calculated the F statistic of the simulated spectra for each image of each object and compared it to the F statistic from the real data. Figure~\ref{fig:fdist} shows the distribution of the F statistics from the simulated spectra, the F statistic from the real data, and the analytical F distribution for image A of \QJ\ as an example. We then calculated the significance of the real data's F statistics by calculating the percentage of the simulated spectra having an F statistic greater than or equal to that of the real data. These calibrated F test significances from the Monte Carlo simulations are given in Tables~\ref{tab:spec_fit_results}--\ref{tab:spec_fit_results_2}. In these tables, we see that the differences between the Monte-Carlo and analytical significances are small for most of the images. We also note that the differences are much larger for the low significance emission lines.
We also simulated the line detection threshold for each spectrum of the combined data by simulating spectra of different EW values and finding those having $1 \sigma$ detections.

\begin{figure}
\epsscale{1.}
\plotone{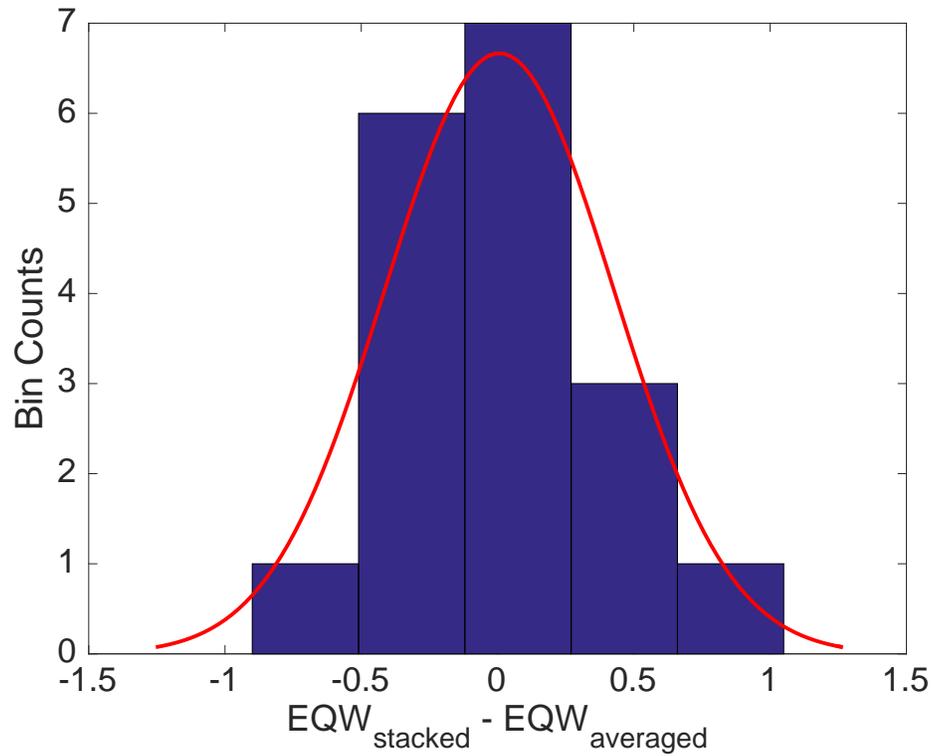}\caption{Histogram of the difference between stacked simulated spectra EWs and averaged simulated spectra EWs. Red curve is the gaussian fit to the histogram. Fit gives a standard deviation of 0.42\,keV and a mean of 0.0065\,keV. Since the peak is very near to zero and the standard deviation is similar to the observed 1$\sigma$ errors for the EW, this means that stacking the spectra is comparable to fitting each individual observation and averaging the results. 
}
\label{fig:EW_sim}
\end{figure}

We also generated simulated spectra to test if the stacked EW measurements are comparable to averaged individual EW measurements, since the line is observed to be a variable between individual observations of gravitational lenses \citep{chartas2017}. 
Figure~\ref{fig:EW_sim} compares these two EWs giving a mean difference of 0.0065\,keV with a standard deviation of 0.42\,keV which is only slightly larger than the 1$\sigma$ uncertainties of the stacked EW measurements. These results show that the stacking process does not bias the mean of the EW from a sequence of observations. 

\begin{figure}
\epsscale{1.}
\plotone{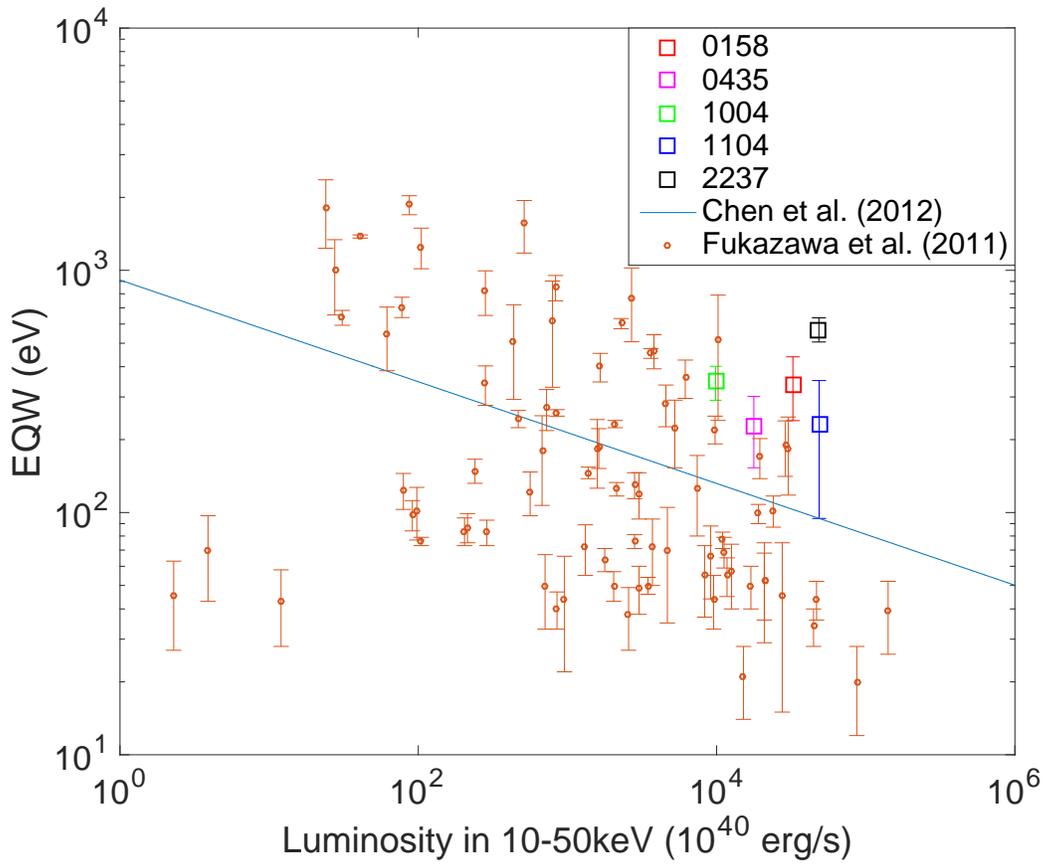}\caption{EW versus Luminosity for the total image spectra of all five lenses using the Combined Data Set. Squares are the data points from this analysis. The blue line is a fit from \citet{chen2012} to the unlensed Seyfert galaxies (orange circles) observed by \emph{Suzaku} in \citet{fukazawa2011}.  
\label{fig:LvsEW}}
\end{figure}

\begin{figure}
\epsscale{1.}
\plotone{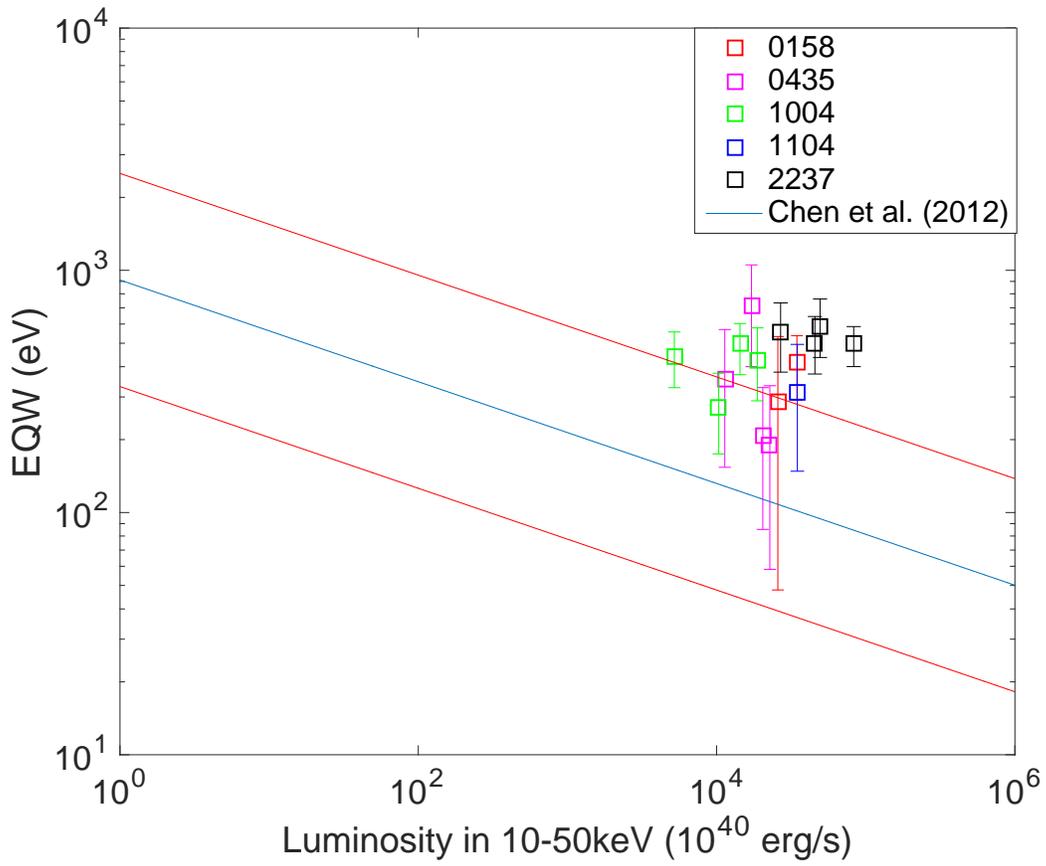}\caption{EW versus Luminosity for the individual images of all five lenses using the Combined Data Set. Squares are the data points from this analysis. The blue and red lines show the mean and scatter of the fit from \citet{chen2012} to the unlensed Seyfert galaxies observed by \emph{Suzaku} in \citet{fukazawa2011} and shown in Figure~\ref{fig:LvsEW}. 
\label{fig:LvsEW_images}}
\end{figure}

\begin{figure}
\epsscale{1.}
\plottwo{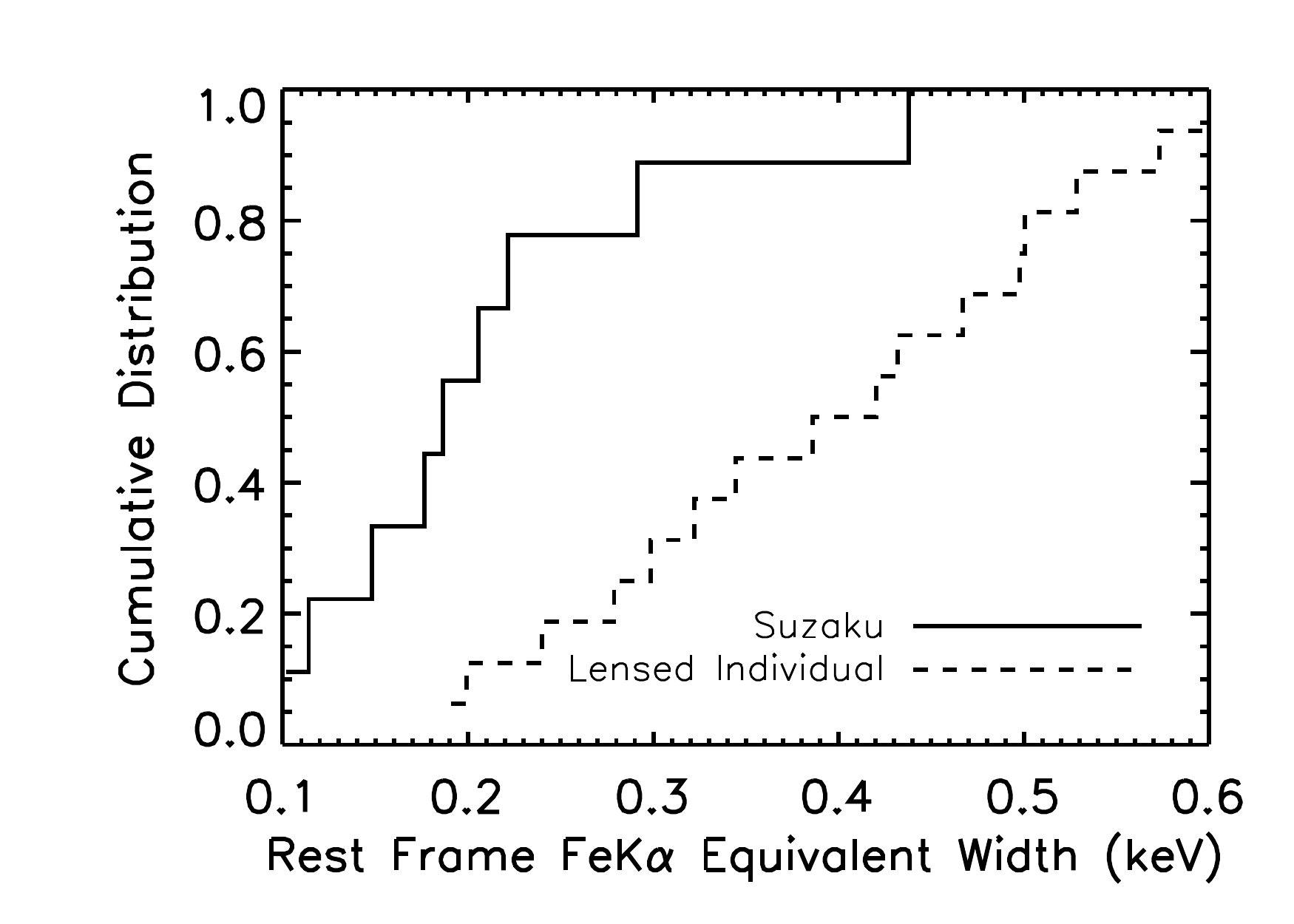}{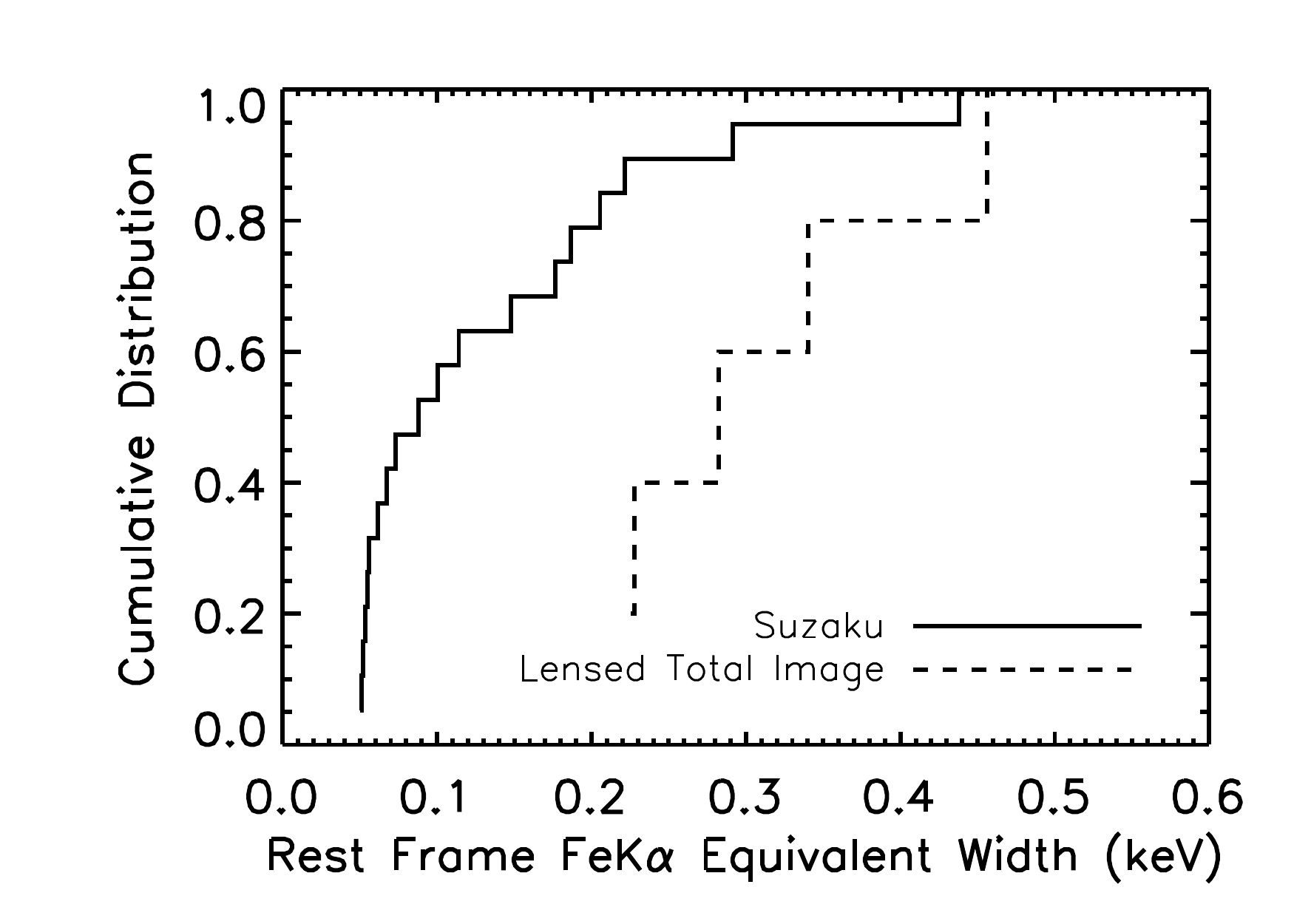}
\caption{Cumulative \feka\ line EW distributions between \emph{Suzaku} and the lensed sample of individual images (left) and \emph{Suzaku} and the lensed sample of total images (right).  The \emph{Suzaku} sample is selected within the luminosity range of the lensed sample and with EW larger than the median detection threshold of the lensed sample.  K-S test results show that the \emph{Suzaku} and the lensed sample of individual images differ by 99.2\% and the \emph{Suzaku} and the lensed sample of total images differ by 99.9\%.
\label{fig:ks}}
\end{figure}

The line energies are measured to be within the range of 5.8 to 6.8\,keV, and the widths of the lines are mostly 1--2$\sigma$ broad compared to the measurement uncertainties. For \Q, the line widths are measured from 0.41 to 0.66 keV with 4--8$\sigma$ broad, confirming the broad line nature first claimed in \citet{dai2003}.  The energy range and broadness of the lines are consistent with \feka\ emission originated from a few $r_g$ around spinning black holes and viewed at different inclination angles, and those 1--2$\sigma$ broad line width measurements are due to the poor signal-to-noise ratios.  
We plot the rest-frame EW and lensing corrected X-ray luminosity of our sample for the five lenses using the results in Table~\ref{tab:spec_fit_results} for the Total and individual images (Figures~\ref{fig:LvsEW}--\ref{fig:LvsEW_images})
and compare with the Iwasawa-Taniguchi or X-ray Baldwin effect, an inverse correlation between the \(EW\) of metal emission lines and the X-ray Luminosity. 
The relation was first discovered by \citet{iwasawa93} for neutral FeK\(\alpha\) lines and the 2--10 keV X-ray luminosity. \citet{fukazawa2011} later showed that the trend holds by including the ionized FeK\(\alpha\) lines and at 10--50 keV X-ray luminosity as well using 88 nearby Seyfert galaxies observed by \emph{Suzaku}. We adopt the fit from ~\citet{chen2012} to the sample of \citet{fukazawa2011} as
\begin{equation}\label{eqn:chenfit}
  \log{\frac{EW_{model}}{\rm eV}} = (2.96 \pm 0.22)-(0.21 \pm 0.07) \log{\frac{L_{X}}{10^{40}\lumin}} \pm (0.44 \pm 0.11).
\end{equation}
The lensed sample shows a positive EW offset from the unlensed systems (Figures~\ref{fig:LvsEW}--\ref{fig:LvsEW_images}). 
For our targets, the predicted rest-frame EW are between 0.08 to 0.15~keV with the mean at 0.11~keV, while the measured values have a range between 0.2--0.7~keV with the mean at 0.42~keV.
We performed the Kolmogorov-Smirnov test between our lensed sample and the \emph{Suzaku} sample \citep{fukazawa2011}.  For the \emph{Suzaku} sample, we selected objects within the luminosity range of the lensed sample between $43.7 < \log{L_X (\lumin)} < 45.0$ and EW values greater than the median detection threshold of the lensed sample, 0.1~keV for individual images and 0.05~keV for total images.  For the lensed sample, the EW of non-detections are set to be half of the detection threshold values.  The K-S test results show that the probability of the null hypothesis that the EW of \emph{Suzaku} sample and the lensed individual image sample are from the same parent distribution is 0.008, and between \emph{Suzaku} sample and the lensed total image sample, the null probability is 0.001. The cumulative EW distributions are shown in Figure~\ref{fig:ks}. The Student's T-test yields similar results with the null probabilities further reduced by a factor of two.  Based on these statistical test results, we concluded that the EW of lensed sample and unlensed sample are different. 

Since our sample of AGN are at high redshifts $z \sim 2$, it is possible that the properties of the \feka\ line are different from the local sample.  However, several studies on high redshift non-lensed samples show little evolution of the EW of \feka\ from the local sample, such as the stacking analyses of \chandra\ deep field sources \citep{falocco2012, falocco2013} with the average EW of 0.07 and 0.14~keV reported.  The \feka\ line of the two brightest sources in the field were measured to have EW of 0.2~keV \citep{iwasawa2015}.
Here, we attribute this offset as a microlensing effect, because microlensing signals are stronger for smaller sources leading to the conclusion that the reflection region is more compact than the X-ray corona. 
Although there are low magnification regions in the magnification patterns, the current sample focuses on lenses with on-going microlensing activities and those non-active lenses usually are not well observed.  

\begin{deluxetable}{lccccccccccccc}
\rotate
    \tabletypesize{\scriptsize}
	\caption{\newline Spectra Fitting Results for the Combined Data Sets \label{tab:spec_fit_results}}
    \tablewidth{0pt}
    \tablehead{
      \colhead{} & \colhead{} & \colhead{} & \colhead{Lens} & \multicolumn{3}{c}{\uline{\feka\ Line Properties}} & \colhead{} & \colhead{} & \colhead{Unlensed} & \multicolumn{2}{c}{\uline{Significance}} & \colhead{\feka\ Det.}\\
    \colhead{Object} & \colhead{Image} & \colhead{$\Gamma$} & \colhead{\(N_H\)} &
    \colhead{\(E_{line}\)} & \colhead{\(\sigma_{line}\)} & \colhead{\(EW_{rest}\)} & 
    \colhead{\(\chi^2_{red}(\nu)\)} & \colhead{\(P(\chi^2/\nu)\)} &
      \colhead{10--50 keV \(L\)\(_{X}\)\tablenotemark{a}} & \colhead{Analy-} & \colhead{Monte} & \colhead{Threshold} \\
      \colhead{} & \colhead{} & \colhead{} & \colhead{($10^{20}cm^{-2}$)} & \colhead{(keV)} & \colhead{(keV)} & \colhead{(keV)} & \colhead{} & \colhead{} & \colhead{($10^{44}\,erg/s$)} & \colhead{tical} & \colhead{Carlo} & \colhead{(keV)}
    }
    \startdata
    0158 & A & \(1.93^{+0.03}_{-0.03}\) & \(0.00^{+0.01}_{-0.00}\) & \(6.49^{+0.07}_{-0.06}\) & \(0.14^{+0.08}_{-0.11}\) & \(0.42^{+0.12}_{-0.12}\) & 1.05(116) & 0.35 & 3.45 & 0.9969 & 0.9988 & 0.1 \\ 
    0158 & B &  \(1.84^{+0.13}_{-0.11}\) & \(0.02^{+0.05}_{-0.02}\)  & \(6.34^{+0.27}_{-0.19}\) & \(<0.49\) & \(0.29^{+0.25}_{-0.24}\) & 0.69(54) & 0.96 & 2.57 & 0.6416 & 0.7419 & 0.3 \\ 
    0158 & Total & \(1.92^{+0.03}_{-0.03}\) & \(0.000^{+0.003}_{-0.000}\) & \(6.50^{+0.07}_{-0.07}\) & \(0.17^{+0.09}_{-0.08}\) & \(0.33^{+0.10}_{-0.09}\) & 0.92(132) & 0.73 & 3.25 & 0.9994 & 0.9996 & 0.1 \\
         \hline
         0435 & A & \(1.93^{+0.06}_{0-0.06}\) & \(0.02^{+0.03}_{-0.02}\) & \(5.98^{+0.12}_{-0.11}\) & \(0.11^{+0.11}_{-0.11}\) & \(0.21^{+0.12}_{-0.12}\) & 0.80(102) & 0.93 & 2.04 & 0.8902 & 0.9023 & 0.1 \\ 
         0435 & B & \(1.89^{+0.5}_{-0.5}\) & \(0.00^{+0.01}_{0.00}\) & \(5.99^{+0.76}_{-0.14}\) & \(0.35^{+0.82}_{-0.21}\) & \(0.36^{+0.21}_{-0.20}\) & 0.90(73) & 0.71 & 1.14 & 0.7891 & 0.8302 & 0.25 \\ 
         0435 & C & \(1.84^{+0.05}_{-0.05}\) & \(0.00^{+0.01}_{-0.00}\) & \(6.41^{+0.43}_{-0.44}\) & \(0.91^{+0.50}_{-0.44}\) & \(0.72^{+0.33}_{-0.32}\) & 0.61(90) & 0.999 & 1.71 & 0.9946 & 0.9983 & 0.07 \\
         0435 & D & \(1.80^{+0.05}_{-0.05}\) & \(0.00^{+0.01}_{-0.00}\) & \(6.28^{+0.14}_{-0.21}\) & \(<0.42\) & \(0.19^{+0.14}_{-0.13}\) & 1.16(76) & 0.17 & 2.27 & 0.4988 & 0.5822 & 0.25 \\ 
         0435 & Total & \(1.88^{+0.02}_{-0.02}\) & \(0.000^{+0.004}_{-0.000}\) & \(6.15^{+0.13}_{-0.12}\) & \(0.28^{+0.10}_{-0.11}\) & \(0.23^{+0.08}_{-0.07}\) & 1.01(142) & 0.47 & 1.78 & 0.9952 & 0.9974 & 0.05 \\
         \hline
         J1004 & A & \(1.72^{+0.04}_{-0.04}\) & \(0.00^{+0.01}_{-0.00}\) & \(6.32^{+0.05}_{-0.10}\) & \(<0.23\) & \(0.44^{+0.12}_{-0.11}\) & 0.97(92) & 0.55 & 0.52 & 0.9993 & 0.9999 & 0.08 \\ 
         J1004 & B & \(1.86^{+0.04}_{-0.04}\) & \(0.01^{+0.02}_{-0.01}\) & \(6.46^{+0.09}_{-0.08}\) & \(0.14^{+0.11}_{-0.08}\) & \(0.27^{+0.10}_{-0.10}\) & 0.86(141) & 0.88 & 1.03 & 0.9941 & 0.9977 & 0.08 \\ 
         J1004 & C & \(1.87^{+0.05}_{-0.05}\) & \(0.03^{+0.3}_{-0.3}\) & \(6.32^{+0.06}_{-0.08}\) & \(0.15^{+0.10}_{-0.09}\) & \(0.50^{+0.11}_{-0.12}\) & 0.82(122) & 0.92 & 1.43 & 0.999992 & 0.9999 & 0.1\\ 
         J1004 & D & \(1.78^{+0.07}_{-0.06}\) & \(0.02^{+0.04}_{-0.02}\) & \(6.30^{+0.07}_{-0.07}\) & \(<0.15\) & \(0.42^{+0.15}_{-0.14}\) & 1.10(86) & 0.24 & 1.87 & 0.9902 & 0.9965 & 0.1 \\ 
         J1004 & Total & \(1.81^{+0.02}_{-0.02}\) & \(0.01^{+0.01}_{-0.01}\) & \(6.37^{+0.03}_{-0.03}\) & \(0.10^{+0.07}_{-0.08}\) & \(0.35^{+0.06}_{-0.06}\) & 0.92(142) & 0.73 & 0.99 & $1-7\times10^{-11}$ & 1.0000 & 0.05\\
         \hline
         1104 & A & \(1.76^{+0.04}_{-0.04}\) & \(0.00^{+0.01}_{-0.00}\) & \(6.78^{+0.30}_{-0.34}\) & \(<0.55\) & \(0.31^{+0.18}_{-0.16}\) & 0.93(84) & 0.65 & 3.48 & 0.8912 & 0.9176 & 0.17 \\ 
         1104 & B & \(1.83^{+0.05}_{-0.05}\) & \(0.00^{+0.01}_{-0.00}\) & ... & ... & ... & 0.79(76) & 0.91 & 9.17 & ... & ... & 0.2\\ 
         1104 & Total & \(1.79^{+0.03}_{-0.03}\) & \(0.000^{+0.004}_{-0.000}\) & \(6.84^{+0.28}_{-0.47}\) & \(0.37^{+0.24}_{-0.13}\) & \(0.23^{+0.12}_{-0.14}\) & 1.09(108) & 0.24 & 4.87 & 0.8173 & 0.8309 & 0.13\\
         \hline
         2237 & A & \(1.84^{+0.03}_{-0.03}\) & \(0.08^{+0.01}_{-0.01}\) & \(6.17^{+0.12}_{-0.12}\) & \(0.60^{+0.13}_{-0.11}\) & \(0.50^{+0.09}_{-0.10}\) & 1.29(120) & 0.02 & 8.30 & $1-4\times10^{-6}$ & 1.0000 & 0.04\\ 
         2237 & B & \(1.84^{+0.06}_{-0.06}\) & \(0.09^{+0.02}_{-0.02}\) & \(6.07^{+0.19}_{-0.20}\) & \(0.53^{+0.16}_{-0.12}\) & \(0.56^{+0.18}_{-0.18}\) & 1.19(86) & 0.11 & 2.69 & 0.9943             & 0.9984 & 0.05\\ 
         2237 & C & \(1.86^{+0.06}_{-0.06}\) & \(0.09^{+0.02}_{-0.02}\) & \(5.76^{+0.12}_{-0.11}\) & \(0.41^{+0.16}_{-0.11}\) & \(0.59^{+0.17}_{-0.15}\) & 0.95(81) & 0.61 & 4.94 & 0.9999 & 0.9998 & 0.05\\ 
         2237 & D & \(1.81^{+0.05}_{-0.05}\) & \(0.10^{+0.02}_{-0.02}\) & \(6.08^{+0.16}_{-0.16}\) & \(0.50^{+0.13}_{-0.11}\) & \(0.50^{+0.14}_{-0.13}\) & 0.93(117) & 0.68 & 4.56 & 0.9998 & 1.0000 & 0.05\\ 
         2237 & Total & \(1.82^{+0.02}_{-0.02}\) & \(0.077^{+0.007}_{-0.007}\) & \(5.89^{+0.08}_{-0.09}\) & \(0.66^{+0.09}_{-0.07}\) & \(0.57^{+0.07}_{-0.06}\) & 1.36(180) & 0.001 & 4.85 & $1-4\times10^{-15}$ & 1.0000 & 0.025\\ 
	\enddata
    \tablecomments{Galactic absorptions are fixed at the values provided in Table~\ref{tab:lensinfo} during the spectral fits.}
    \tablenotetext{a}{Unabsorbed and unlensed luminosity, where the macro magnification corrections are based on Table~\ref{tab:mag_map_props}.}
\end{deluxetable}

\begin{deluxetable}{lcccccccccccc}
\rotate
    \tabletypesize{\scriptsize}
	\caption{\newline Spectra Fitting Results for Data Set 1 \label{tab:spec_fit_results_1}}
    \tablewidth{0pt}
    \tablehead{
    \colhead{} & \colhead{} & \colhead{} & \colhead{Lens} & \multicolumn{3}{c}{\uline{\feka\ Line Properties}} & \colhead{} & \colhead{} & \colhead{Unlensed} & \multicolumn{2}{c}{\uline{Significance}}\\
    \colhead{Object} & \colhead{Image} & \colhead{$\Gamma$} & \colhead{\(N_H\)} &
    \colhead{\(E_{line}\)} & \colhead{\(\sigma_{line}\)} & \colhead{\(EW_{rest}\)} & 
    \colhead{\(\chi^2_{red}(\nu)\)} & \colhead{\(P(\chi^2/\nu)\)} &
    \colhead{10--50 keV \(L\)\(_{X}\)\tablenotemark{a}} & \colhead{Analytical} & \colhead{Monte Carlo}\\
    \colhead{} & \colhead{} & \colhead{} & \colhead{($10^{20}cm^{-2}$)} & \colhead{(keV)} & \colhead{(keV)} & \colhead{(keV)} & \colhead{} & \colhead{} & \colhead{($10^{44}\,erg/s$)} & \colhead{} & \colhead{}
    }
    \startdata
         0158 & A & \(1.91^{+0.08}_{-0.07}\) & \(<0.02\) & \(6.46^{+0.06}_{-0.09}\) & \(<0.22\) & \(0.37^{+0.22}_{-0.24}\) & 1.02(105) & 0.43 & 3.00 & 0.5458 & 0.5534\\ 
         0158 & B &  \(1.72^{+0.18}_{-0.13}\) & \(<0.07\)  & ... & ... & ... & 0.90(34) & 0.63 & 3.35 & ... & ...\\ 
         0158 & Total & \(1.87^{+0.06}_{-0.06}\) & \(<0.01\) & \(6.42^{+0.15}_{-0.15}\) & \(<0.32\) & \(0.36^{+0.23}_{-0.23}\) & 0.97(63) & 0.54 & 3.30 & 0.7415 & 0.7773\\
         \hline
         0435 & A & \(1.97^{+0.17}_{0-0.16}\) & \(0.10^{+0.08}_{-0.07}\) & \(6.53^{+1.51}_{-0.86}\) & \(<0.73\) & \(0.46^{+0.77}_{-0.39}\) & 0.90(65) & 0.70 & 1.35 & 0.6564 & 0.7210\\ 
         0435 & B & \(1.84^{+0.18}_{-0.14}\) & \(<0.12\) & \(6.64^{+0.12}_{-0.19}\) & \(<0.35\) & \(0.90^{+0.63}_{-0.50}\) & 0.90(42) & 0.66 & 1.20 & 0.8315 & 0.8924\\ 
         0435 & C & \(1.51^{+0.11}_{-0.11}\) & \(<0.03\) & \(6.43^{+0.11}_{-0.11}\) & \(<0.32\) & \(0.65^{+0.42}_{-0.39}\) & 1.04(33) & 0.40 & 2.30 & 0.6916 & 0.7609\\
         0435 & D & \(2.01^{+0.23}_{-0.21}\) & \(0.12^{+0.11}_{-0.10}\) & ... & ... & ... & 0.89(47) & 0.73 & 2.30 & ... & ...\\ 
         0435 & Total & \(1.82^{+0.07}_{-0.07}\) & \(<0.06\) & \(6.52^{+0.15}_{-0.13}\) & \(<0.35\) & \(0.35^{+0.18}_{-0.16}\) & 0.99(77) & 0.56 & 1.70 & 0.8983 & 0.9297\\
         \hline
         J1004 & A & \(1.80^{+0.08}_{-0.07}\) & \(<0.06\) & \(6.28^{+0.08}_{-0.10}\) & \(<0.33\) & \(0.63^{+0.21}_{-0.20}\) & 1.15(71) & 0.18 & 0.52 & 0.9841 & 0.9966\\ 
         J1004 & B & \(1.89^{+0.06}_{-0.04}\) & \(<0.03\) & \(6.57^{+0.09}_{-0.08}\) & \(<0.23\) & \(0.50^{+0.17}_{-0.19}\) & 0.92(102) & 0.71 & 0.85 & 0.9897 & 0.9944\\ 
         J1004 & C & \(1.90^{+0.08}_{-0.08}\) & \(0.05^{+0.04}_{-0.04}\) & \(6.22^{+0.29}_{-0.25}\) & \(0.49^{+0.28}_{-0.20}\) & \(0.59^{+0.26}_{-0.26}\) & 0.83(80) & 0.86 & 1.32 & 0.9507 & 0.9739\\ 
         J1004 & D & \(1.74^{+0.11}_{-0.10}\) & \(<0.11\) & \(6.26^{+0.06}_{-0.09}\) & \(<0.20\) & \(0.50^{+0.22}_{-0.19}\) & 1.23(65) & 0.10 & 2.19 & 0.8882 & 0.9276\\ 
         J1004 & Total & \(1.84^{+0.04}_{-0.04}\) & \(0.03^{+0.02}_{-0.02}\) & \(6.37^{+0.82}_{-0.82}\) & \(<0.07\) & \(0.37^{+0.14}_{-0.14}\) & 0.92(129) & 0.73 & 0.96 & $1-4\times10^{-6}$ & 1.0000\\
         \hline
         1104 & A & \(1.77^{+0.05}_{-0.05}\) & \(<0.007\) & \(7.06^{+0.35}_{-0.35}\) & \(<0.82\) & \(0.42^{+0.28}_{-0.26}\) & 0.99(78) & 0.50 & 3.77 & 0.6500 & 0.7031\\ 
         1104 & B & \(1.87^{+0.06}_{-0.06}\) & \(<0.02\) & ... & ... & ... & 0.70(103) & 0.99 & 8.42 & ... & ...\\ 
         1104 & Total & \(1.79^{+0.04}_{-0.04}\) & \(<0.005\) & \(6.99^{+0.27}_{-0.24}\) & \(<0.61\) & \(0.30^{+0.16}_{-0.16}\) & 1.10(103) & 0.23 & 5.04 & 0.7623 & 0.7835\\
         \hline
         2237 & A & \(1.85^{+0.04}_{-0.04}\) & \(0.08^{+0.01}_{-0.01}\) & \(6.46^{+0.07}_{-0.08}\) & \(0.22^{+0.09}_{-0.07}\) & \(0.37^{+0.08}_{-0.08}\) & 1.38(110) & 0.005 & 8.41 & 0.9998 & 1.0000\\ 
         2237 & B & \(1.89^{+0.08}_{-0.08}\) & \(0.09^{+0.02}_{-0.02}\) & \(6.29^{+0.22}_{-0.19}\) & \(0.35^{+0.13}_{-0.14}\) & \(0.45^{+0.21}_{-0.20}\) & 1.29(86) & 0.04 & 2.48 & 0.8984 & 0.9287\\ 
         2237 & C & \(1.96^{+0.06}_{-0.06}\) & \(0.08^{+0.03}_{-0.02}\) & \(6.04^{+0.14}_{-0.13}\) & \(0.43^{+0.11}_{-0.09}\) & \(0.85^{+0.25}_{-0.24}\) & 1.13(91) & 0.19 & 3.36 & 0.9972 & 0.9988\\ 
         2237 & D & \(1.87^{+0.06}_{-0.06}\) & \(0.13^{+0.02}_{-0.02}\) & \(6.12^{+0.21}_{-0.21}\) & \(0.54^{+0.18}_{-0.15}\) & \(0.53^{+0.18}_{-0.18}\) & 0.93(92) & 0.67 & 4.83 & 0.9971 & 0.9996\\ 
         2237 & Total & \(1.88^{+0.03}_{-0.03}\) & \(0.082^{+0.008}_{-0.008}\) & \(5.99^{+0.10}_{-0.11}\) & \(0.60^{+0.09}_{-0.08}\) & \(0.56^{+0.09}_{-0.08}\) & 1.29(148) & 0.009 & 4.64 & $1-10^{-10}$ & 1.0000\\ 
	\enddata
        \tablecomments{Galactic absorptions are fixed at the values provided in Table~\ref{tab:lensinfo} during the spectral fits.}
    \tablenotetext{a}{Unabsorbed and unlensed luminosity, where the macro magnification corrections are based on Table~\ref{tab:mag_map_props}.}
\end{deluxetable}

\begin{deluxetable}{lcccccccccccc}
\rotate
    \tabletypesize{\scriptsize}
	\caption{\newline Spectra Fitting Results for Data Set 2 \label{tab:spec_fit_results_2}}
    \tablewidth{0pt}
    \tablehead{
    \colhead{} & \colhead{} & \colhead{} & \colhead{Lens} & \multicolumn{3}{c}{\uline{\feka\ Line Properties}} & \colhead{} & \colhead{} & \colhead{Unlensed} & \multicolumn{2}{c}{\uline{Significance}}\\
    \colhead{Object} & \colhead{Image} & \colhead{$\Gamma$} & \colhead{\(N_H\)} &
    \colhead{\(E_{line}\)} & \colhead{\(\sigma_{line}\)} & \colhead{\(EW_{rest}\)} & 
    \colhead{\(\chi^2_{red}(\nu)\)} & \colhead{\(P(\chi^2/\nu)\)} &
    \colhead{10--50 keV \(L\)\(_{X}\)\tablenotemark{a}} & \colhead{Analytical} & \colhead{Monte Carlo}\\
    \colhead{} & \colhead{} & \colhead{} & \colhead{($10^{20}cm^{-2}$)} & \colhead{(keV)} & \colhead{(keV)} & \colhead{(keV)} & \colhead{} & \colhead{} & \colhead{($10^{44}\,erg/s$)} & \colhead{} & \colhead{}
    }
    \startdata
         0158 & A & \(1.94^{+0.04}_{-0.04}\) & \(<0.01\) & \(6.50^{+0.11}_{-0.10}\) & \(0.20^{+0.16}_{-0.11}\) & \(0.41^{+0.15}_{-0.15}\) & 0.90(121) & 0.77 & 3.46 & 0.9859 & 0.9971\\ 
         0158 & B &  \(1.90^{+0.15}_{-0.14}\) & \(<0.11\)  & $6.43^{+0.10}_{-0.14}$ & $<0.9$ & $0.34^{+0.22}_{-0.24}$ & 0.95(44) & 0.57 & 2.22 & 0.4912 & 0.7904\\ 
         0158 & Total & \(1.94^{+0.03}_{-0.03}\) & \(<0.005\) & \(6.52^{+0.09}_{-0.08}\) & \(0.32^{+0.09}_{-0.09}\) & \(0.39^{+0.13}_{-0.24}\) & 0.76(132) & 0.98 & 3.09 & 0.9997 & 1.0000\\
         \hline
         0435 & A & \(1.93^{+0.05}_{0-0.04}\) & \(<0.02\) & \(6.05^{+0.07}_{-0.09}\) & \(<0.42\) & \(0.22^{+0.08}_{-0.22}\) & 0.70(104) & 0.99 & 2.06 & 0.9541 & 0.9909\\ 
         0435 & B & \(1.91^{+0.06}_{-0.06}\) & \(<0.01\) & \(6.24^{+0.46}_{-0.11}\) & \(<0.15\) & \(0.25^{+0.12}_{-0.15}\) & 0.97(69) & 0.55 & 1.09 & 0.6860 & 0.8385\\ 
         0435 & C & \(1.88^{+0.05}_{-0.04}\) & \(<0.02\) & \(6.53^{+0.06}_{-0.38}\) & \(<0.13\) & \(0.28^{+0.24}_{-0.17}\) & 0.93(85) & 0.66 & 1.70 & 0.9050 & 0.9779\\
         0435 & D & \(1.80^{+0.06}_{-0.06}\) & \(<0.009\) & $5.97^{+0.35}_{-0.22}$ & $<0.47$ & $0.26^{+0.19}_{-0.20}$ & 1.14(77) & 0.19 & 2.23 & 0.4836 & 0.7069\\ 
         0435 & Total & \(1.91^{+0.02}_{-0.02}\) & \(<0.003\) & \(6.06^{+0.11}_{-0.10}\) & \(0.24^{+0.08}_{-0.08}\) & \(0.25^{+0.08}_{-0.07}\) & 1.08(116) & 0.27 & 1.71 & 0.9942 & 0.9998\\
         \hline
         J1004 & A & \(1.61^{+0.06}_{-0.05}\) & \(<0.02\) & \(6.29^{+0.07}_{-0.07}\) & \(<0.18\) & \(0.40^{+0.20}_{-0.16}\) & 0.94(74) & 0.63 & 0.53 & 0.9551 & 0.9746\\ 
         J1004 & B & \(1.86^{+0.05}_{-0.04}\) & \(<0.03\) & \(6.18^{+0.17}_{-0.16}\) & \(0.26^{+0.15}_{-0.15}\) & \(0.32^{+0.15}_{-0.15}\) & 0.70(102) & 0.99 & 1.12 & 0.9722 & 0.9957\\ 
         J1004 & C & \(1.85^{+0.04}_{-0.04}\) & \(<0.02\) & \(6.34^{+0.07}_{-0.09}\) & \(0.16^{+0.11}_{-0.13}\) & \(0.52^{+0.16}_{-0.14}\) & 0.84(90) & 0.84 & 1.50 & 0.9991 & 0.9995\\ 
         J1004 & D & \(1.83^{+0.10}_{-0.06}\) & \(<0.08\) & \(6.42^{+0.05}_{-0.06}\) & \(<0.12\) & \(0.43^{+0.17}_{-0.19}\) & 0.81(72) & 0.87 & 1.53 & 0.9533 & 0.9774\\ 
         J1004 & Total & \(1.81^{+0.02}_{-0.02}\) & \(<0.01\) & \(6.27^{+0.08}_{-0.09}\) & \(0.29^{+0.13}_{-0.11}\) & \(0.43^{+0.10}_{-0.10}\) & 0.89(131) & 0.80 & 0.96 & $1-2\times10^{-6}$ & 1.0000\\
         \hline
         1104 & A & \(1.75^{+0.08}_{-0.08}\) & \(<0.05\) & \(6.53^{+0.08}_{-0.08}\) & \(<0.19\) & \(0.46^{+0.19}_{-0.17}\) & 0.66(56) & 0.98 & 2.76 & 0.9871 & 0.9941\\ 
         1104 & B & \(1.80^{+0.08}_{-0.08}\) & \(<0.06\) & ... & ... & ... & 0.67(60) & 0.98 & 8.94 & ... & ...\\ 
         1104 & Total & \(1.80^{+0.05}_{-0.05}\) & \(<0.03\) & \(6.44^{+0.61}_{-0.56}\) & \(<0.16\) & \(0.25^{+0.13}_{-0.11}\) & 0.85(64) & 0.80 & 4.19 & 0.9491 & 0.9622\\
         \hline
         2237 & A & \(1.86^{+0.05}_{-0.05}\) & \(0.11^{+0.02}_{-0.02}\) & \(5.97^{+0.19}_{-0.189}\) & \(0.39^{+0.20}_{-0.20}\) & \(0.27^{+0.13}_{-0.12}\) & 0.87(94) & 0.81 & 8.34 & 0.9521 & 0.9674\\ 
         2237 & B & \(1.81^{+0.10}_{-0.10}\) & \(0.06^{+0.03}_{-0.03}\) & \(6.02^{+0.27}_{-0.26}\) & \(0.71^{+0.23}_{-0.19}\) & \(0.82^{+0.37}_{-0.31}\) & 0.97(55) & 0.53 & 3.30 & 0.9788 & 0.9921\\ 
         2237 & C & \(1.82^{+0.09}_{-0.09}\) & \(0.10^{+0.04}_{-0.03}\) & \(5.54^{+0.13}_{-0.14}\) & \(0.38^{+0.17}_{-0.13}\) & \(0.57^{+0.23}_{-0.18}\) & 0.86(76) & 0.81 & 7.25 & 0.9984 & 0.9867\\ 
         2237 & D & \(1.63^{+0.09}_{-0.09}\) & \(0.02^{+0.03}_{-0.02}\) & \(6.16^{+0.23}_{-0.23}\) & \(0.33^{+0.44}_{-0.15}\) & \(0.38^{+0.25}_{-0.21}\) & 1.09(77) & 0.28 & 4.59 & 0.7631 & 0.7845\\ 
         2237 & Total & \(1.78^{+0.13}_{-0.17}\) & \(0.07^{+0.01}_{-0.01}\) & \(5.74^{+0.13}_{-0.17}\) & \(0.69^{+0.25}_{-0.16}\) & \(0.55^{+0.12}_{-0.11}\) & 0.83(136) & 0.92 & 5.49 & $1-4\times10^{-9}$ & 1.0000\\ 
	\enddata
        \tablecomments{Galactic absorptions are fixed at the values provided in Table~\ref{tab:lensinfo} during the spectral fits.}
    \tablenotetext{a}{Unabsorbed and unlensed luminosity, where the macro magnification corrections are based on Table~\ref{tab:mag_map_props}.}
\end{deluxetable}

\section{Microlensing Analysis}
\label{sec:ML}
We performed a microlensing analysis to interpret the positive EW offset measured in lensed quasars.
In magnitude units, we have
\begin{equation}\label{eqn:EW_ml}
f = -2.5\log_{10}{\frac{EW_{data}}{EW_{fit}}},
\end{equation}
where $f$ is the differential magnification magnitude between the magnification of the reflection region and the corona.
Here, we have included a modest evolution effect, assuming that the rest-frame EW of high redshift quasars are higher than the local ones to be $EW_{fit} = 0.2$~keV \citep{iwasawa2015}.
We generated magnification maps of these five lenses using the inverse polygon mapping algorithm \citep[]{mediavilla2006, mediavilla2011b}. These magnification maps are 16,000$^2$ pixels and each pixel has a length scale of 0.685 $r_g$ for the five lenses. The rest of the pertinent magnification map properties are listed in Table~\ref{tab:mag_map_props} and further details on the magnification maps can be found in \citet{guer2017}. We then generated images of ``average'' AGN corona and the reflection region that are modified by the relativistic effects caused by the black hole using the software \verb+KERTAP+ \citep{chen2013a, chen2013b, chen2015}. 

\begin{figure}
\epsscale{1.}
\plotone{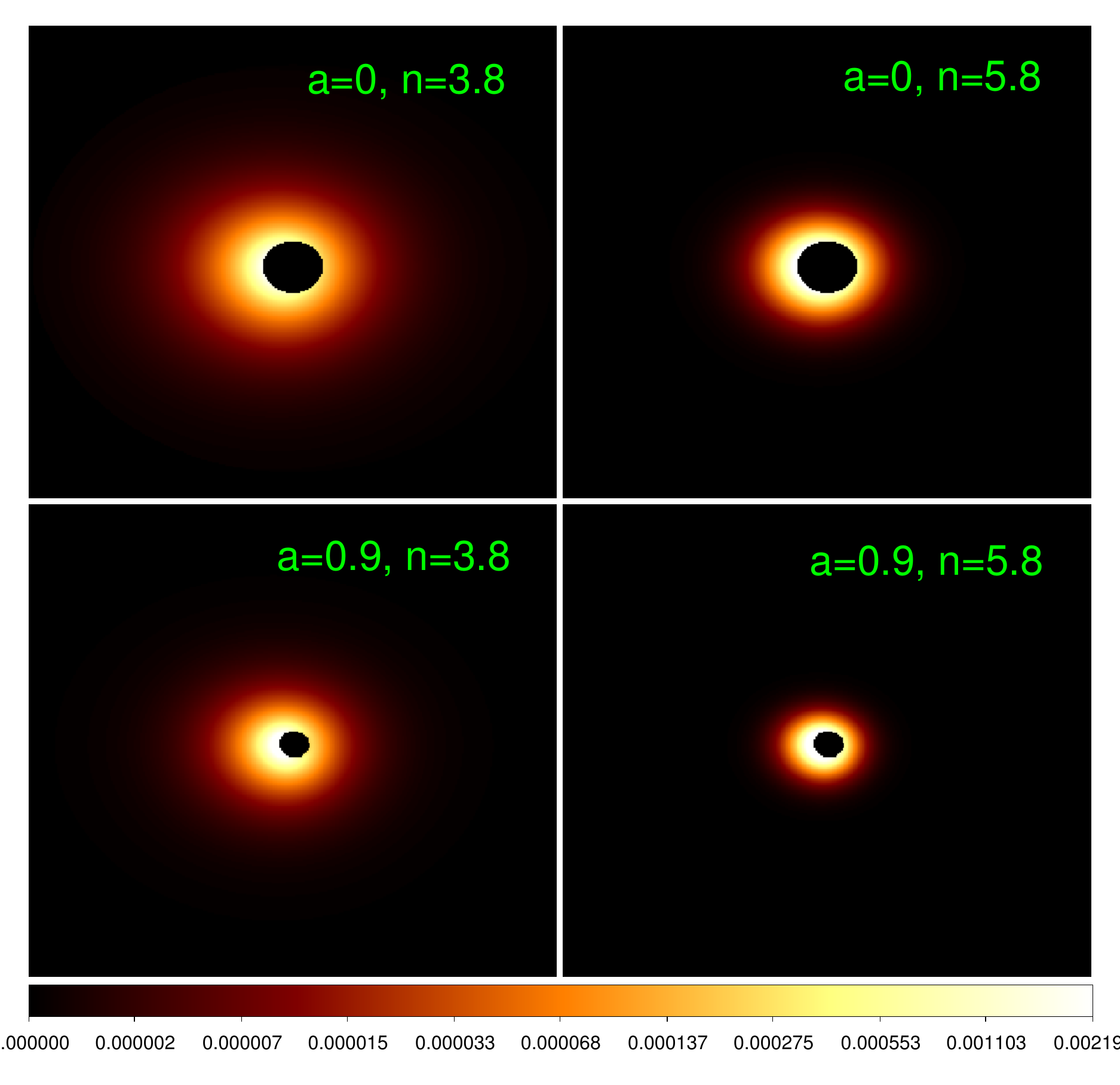}
\caption{Example Kerr images of the \feka\ emissivity profiles from KERTAP \citep{chen2015} with $a=0, n=3.8$ (top, left), $a=0, n=5.8$ (top, right), $a=0.9, n=3.8$ (bottom, left), and $a=0.9, n=5.8$ (bottom, right), where the inclination angle is fixed at 40$^\circ$. 
}
\label{fig:kerr}
\end{figure}

We assumed that the X-ray corona and reflection region are located very close to the disk in Keplerian motion, following power-law emissivity profiles, \(I \propto r^{-n}\), but with different emissivity indices. The models have three parameters: the Kerr spin parameter $a$ of the black hole, the power-law index for the emissivity profile of the reflection region $n$, and the inclination angle of the accretion disk. To simplify the analysis, we set the inclination angle to 40 degrees, a typical inclination angle for Type I AGN. The emissivity index $n$ for the reflection region was varied from 3.0 to 6.2 in steps of \(0.4\) and the spin $a$ was varied between 0 and 0.998 in steps of 0.1. Some example Kerr images of the emissivity profiles are shown in Figure~\ref{fig:kerr}.
These resulting images were then convolved with the magnification maps of each lens image to estimate the amount of microlensing that the corona and reflection region would experience at different locations of the magnification maps. We performed these convolutions in flux units and then converted to magnitude scales.
We also tested to see if the orientation of the corona and reflection region with respect to the magnification map matters by rotating the images by 90 degrees, and in general this will not invoke a significant change in the parameter estimations discussed below.
The half light radius for the X-ray continuum emission is expected to be $\sim$10\,\(r_g\) (gravitational radii) \citep{dai2010, mosquera2013}, corresponding to emissivity indices of \(n=2.2\) -- 3.4 for different spins.  We then subtracted the two convolved magnification maps from the continuum and reflection models to estimate the differential microlensing between the two emission regions, \(f_j=\mu_{con} - \mu_{ref}\) in magnitude units, as a function of source position on the magnification map for a lensed image and common values of $a$ and $n$. 
We obtained distributions of these subtracted convolutions by making histograms of the values for randomly selected points in the subtracted convolved images. 

\begin{deluxetable}{lcccccccccccc}
    \tabletypesize{\scriptsize}
	\caption{\newline Lensing Parameters of the Sample \label{tab:mag_map_props}}
    \tablewidth{0pt}
    \tablehead{
    \colhead{Object} & \colhead{Image} & \colhead{$R_E$\tablenotemark{a}} & \colhead{$\kappa$} & \colhead{$\gamma$} & \colhead{Macro} & \colhead{Black Hole Mass\tablenotemark{b}} \\
    \colhead{} & \colhead{} & \colhead{(light-days)} & \colhead{} & \colhead{} & \colhead{Magnification} & \colhead{($10^9$ $M_\odot$)} 
    }
    \startdata
    0158 & A & 13.2 & 0.348 & 0.428 & 4.13 & 0.16 (MgII) \\
    0158 & B & 13.2 & 0.693 & 0.774 & 1.98 & 0.16 (MgII) \\
    0435 & A & 11.4 & 0.445 & 0.383 & 6.20 & 0.50 (CIV) \\
    0435 & B & 11.4 & 0.539 & 0.602 & 6.67 & 0.50 (CIV) \\
    0435 & C & 11.4 & 0.444 & 0.396 & 6.57 & 0.50 (CIV) \\
    0435 & D & 11.4 & 0.587 & 0.648 & 4.01 & 0.50 (CIV) \\
    J1004 & A & 9.1 & 0.763 & 0.300 & 29.6 & 0.39 (MgII) \\
    J1004 & B & 9.1 & 0.696 & 0.204 & 19.7 & 0.39 (MgII) \\
    J1004 & C & 9.1 & 0.635 & 0.218 & 11.7 & 0.39 (MgII) \\
    J1004 & D & 9.1 & 0.943 & 0.421 & 5.75 & 0.39 (MgII) \\
    1104 & A & 9.1 & 0.610 & 0.512 & 9.09 & 0.59 ($H_\beta$) \\
    1104 & B & 9.1 & 0.321 & 0.217 & 2.42 & 0.59 ($H_\beta$) \\
    2237 & A & 38.5 & 0.390 & 0.400 & 4.71 & 1.20 ($H_\beta$) \\
    2237 & B & 38.5 & 0.380 & 0.390 & 4.30 & 1.20 ($H_\beta$) \\
    2237 & C & 38.5 & 0.740 & 0.730 & 2.15 & 1.20 ($H_\beta$) \\
    2237 & D & 38.5 & 0.640 & 0.620 & 3.92 & 1.20 ($H_\beta$) \\
	\enddata
    \tablenotetext{a}{Einstein radius size on the source plane for a 0.3 $M_\odot$ star.}
    \tablenotetext{b}{The balck hole mass and the emission line used for the estimates by~\citet{morgan2010} (\QJ, \HEone, \SD) and~\citet{assef2011} (\HEtwo, \Q).}
\end{deluxetable}

\begin{figure}
\epsscale{1.}
\plottwo{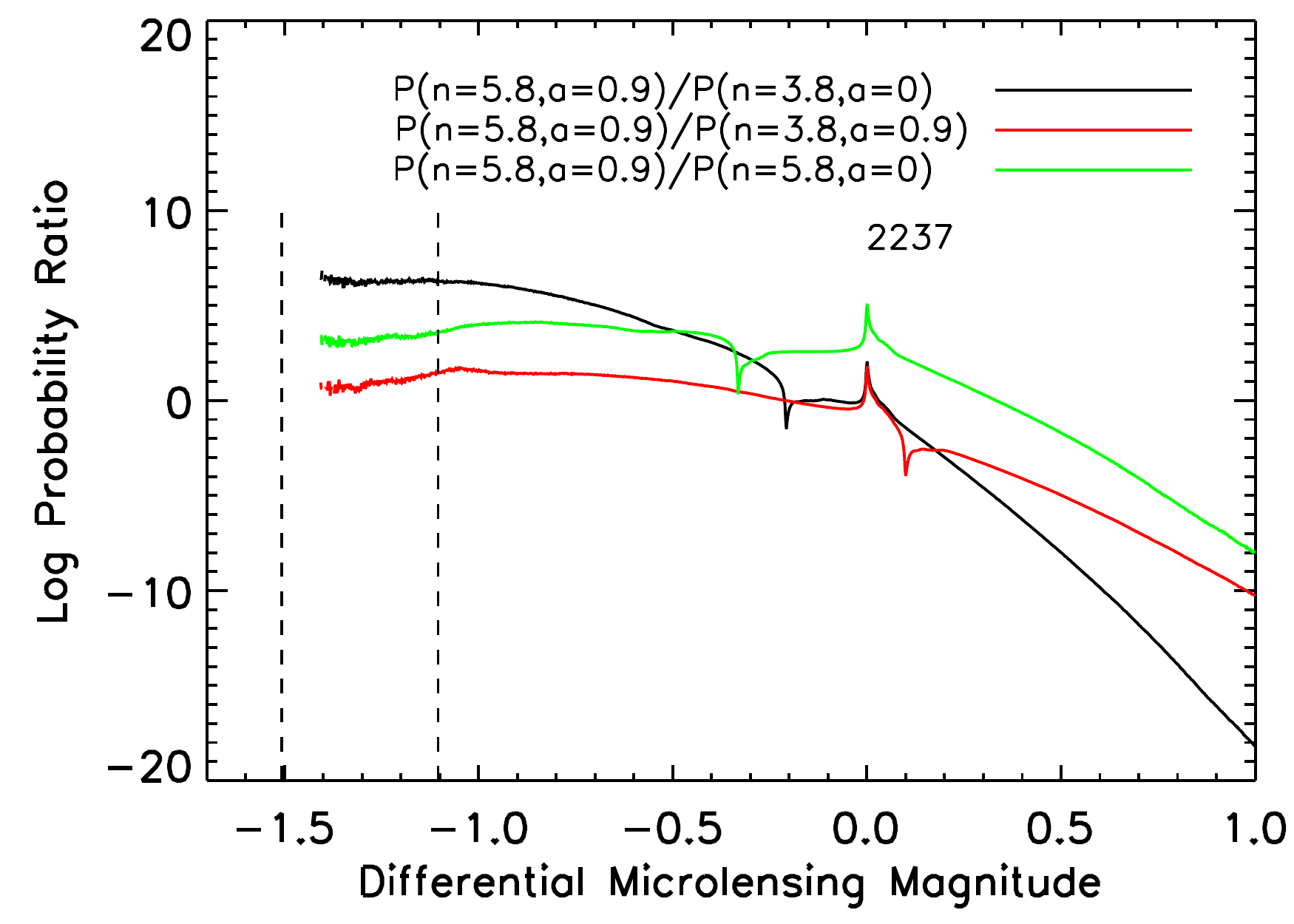}{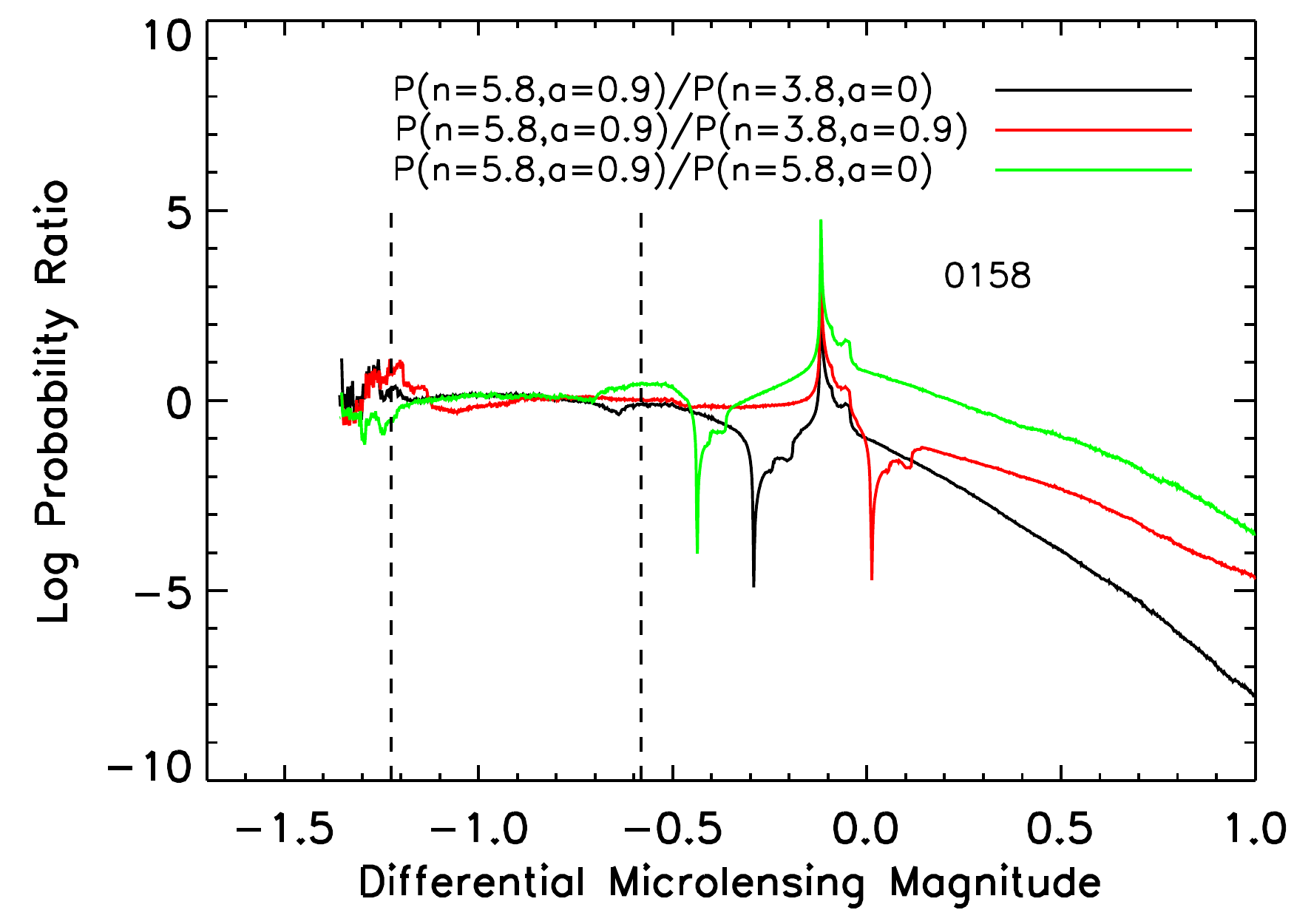}
\caption{Sample model likelihood ratios as a function of the differential microlensing magnitude for \Q\,A (left) and \QJ\,A (right). The logarithm  likelihood ratios are between the model of ($n=5.8$, $a=0.9$) and the models of ($n=3.8$, $a=0$, black), ($n=3.8$, $a=0.9$, red) and ($n=5.8$, $a=0.0$, green).  The vertical dashed lines indicate the observed differential microlensing magnitude range including uncertainties. 
\label{fig:mlMAGdist}}
\end{figure}

We then performed a likelihood analysis using the microlensing magnification distributions and the \(EW\)s we measured in the data. For a given image and fixed $n$ and $a$, the likelihood of the data given the model is
\begin{equation}\label{eqn:L(n)_image}
L_{image}(n,a)=A\sum_j b_j(n,a) e^{-\chi_j^2[f_j(n,a)]/2},
\end{equation}
where \(A\) is a normalization constant and $b_j$ is the bin height of the $j$th value from the convolved histograms discussed previously. The chi-square is calculated from
\begin{equation}\label{eqn:chi(n)}
\chi_j^2[f_j(n,a)]=\left(\frac{f_j(n,a)+2.5 \log_{10}{EW_{data}}-2.5 \log_{10}{EW_{fit}}}{2.5 EW_{err}/(EW_{data}\ln{10})}\right)^2,
\end{equation}
where \(EW_{data}\) and \(EW_{err}\) are respectively the \(EW\) and the \(EW\) uncertainty from the spectra analysis done in Section~\ref{sec:spec} and $f_j$ is the amount of differential microlensing between the continuum and reflection regions in magnitudes. 
Using Equations~\ref{eqn:chenfit}--\ref{eqn:chi(n)}, we then have a likelihood as a function of $n$ and $a$, the index of the emissivity profile and spin parameter, for each image of each object.
Examples for \Q\ A and \QJ\ A are shown in Figure~\ref{fig:mlMAGdist}, where we plot the model likelihood ratios compare to the $a=0.9$ and $n=5.8$ model.
We then combined the likelihoods of all the images from a target
\begin{equation}\label{eqn:L(n)_total}
L_{total}(n, a)=\prod_{image}B L_{image}(n, a),
\end{equation}
where B is the new normalization constant and the multiplication applies to all the images of the target. 
After calculating $L_{total}(n, a)$ for a grid of $(n, a)$ combinations, we can then marginalize over either $a$ or $n$ to obtain the posterior probability for the emissivity index or spin, separately. 
We have discarded \HEtwo B, the non-detection case, in the microlensing analysis, and since it contributes to less than a tenth of the sample, we do not expect that our results will change significantly.

\begin{figure}
\epsscale{1.}
\plottwo{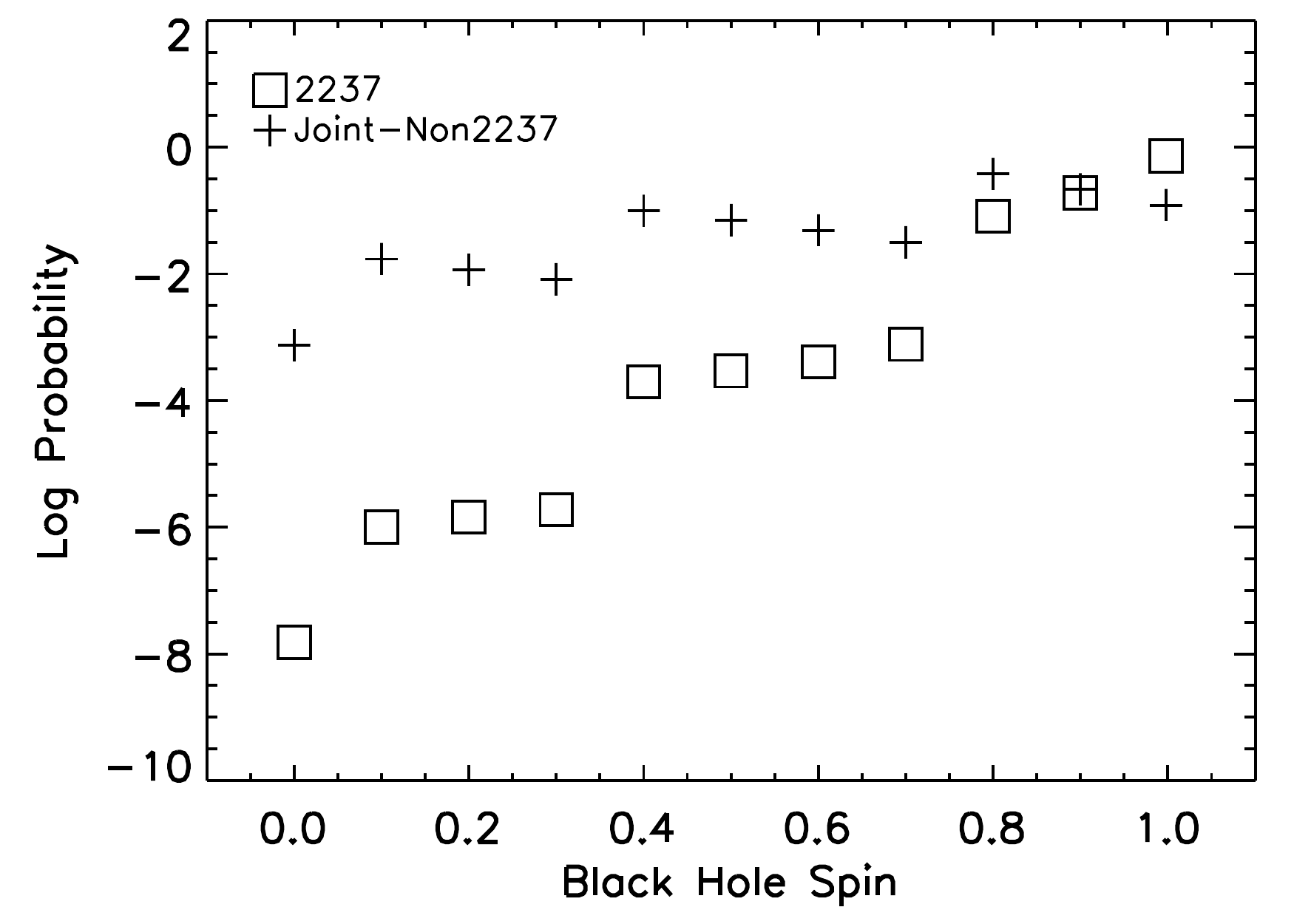}{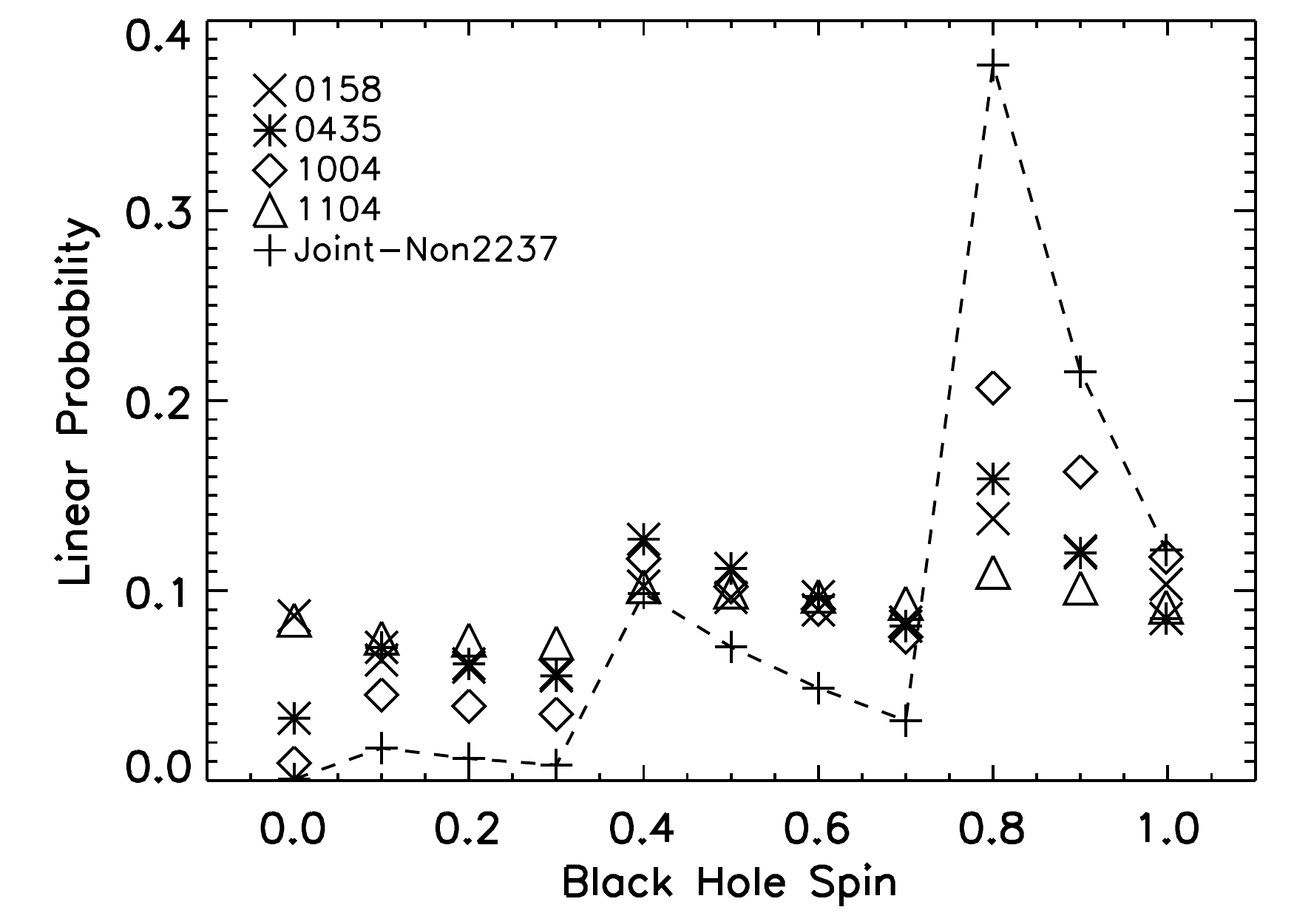}
\caption{(left) Logarithm likelihood as a function of black hole spin for \Q\ and the joint sample excluding \Q. (right) Linear likelihood as a function of black hole spin for \QJ, \HEone, \SD, \HEtwo\ and the joint sample of the four targets.
The 68\% and 90\% confidence limits for \Q\ are $a > 0.92$ and $a>0.83$, and the corresponding limits for the joint sample are $a=0.8\pm0.16$ and $a > 0.41$.
}
\label{fig:L(a)_total}
\end{figure}

\begin{figure}
\epsscale{1.}
\plottwo{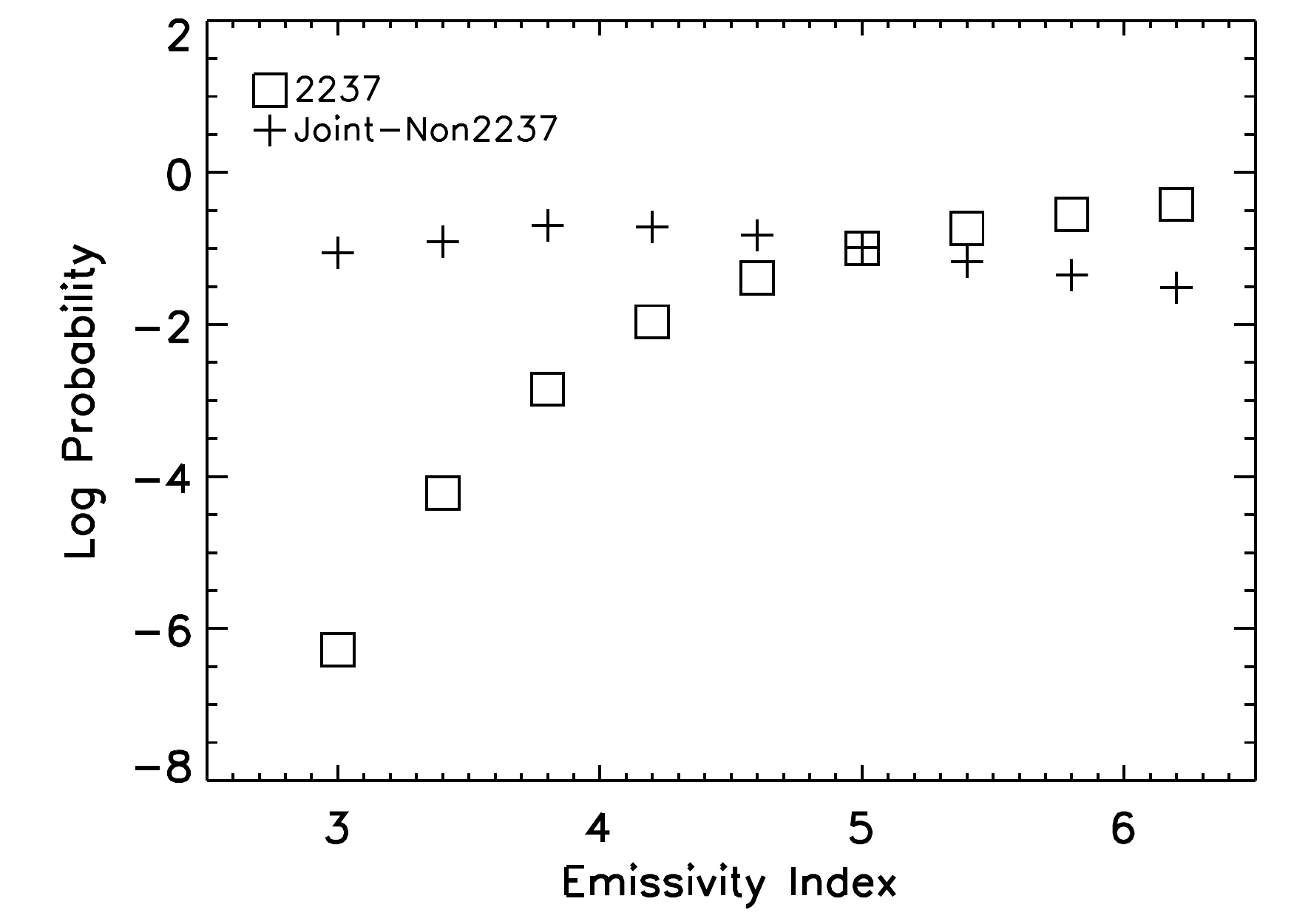}{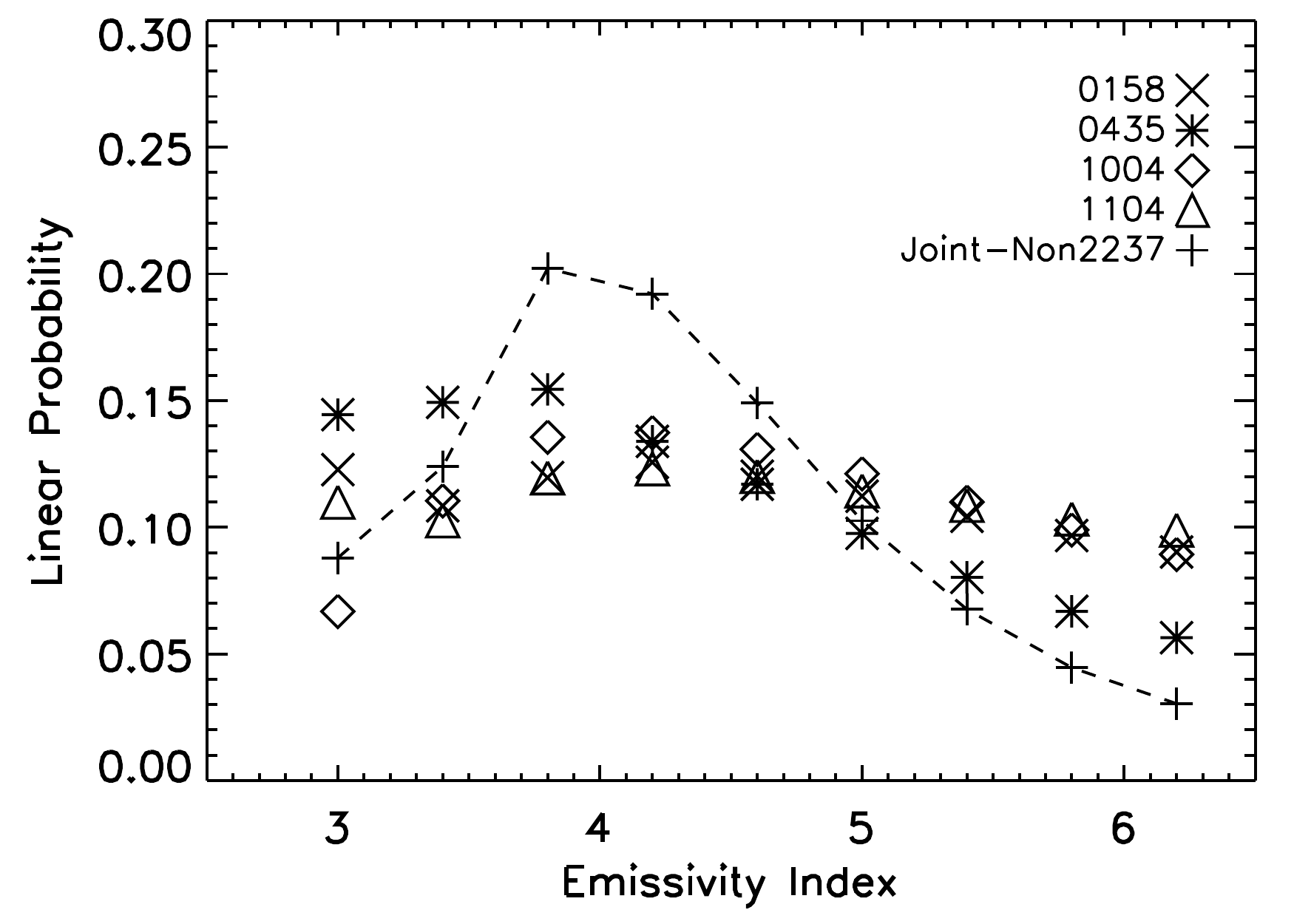}
\caption{(left) Logarithm likelihood as a function of emissivity index for \Q\ and the joint sample excluding \Q. (right) Linear likelihood  for \QJ, \HEone, \SD, \HEtwo\ and the joint sample of the four targets.
The 68\% and 90\% confidence limits for \Q\ are $n > 5.4$ and $n > 4.9$, and the corresponding limits for the joint sample are 
 $n=4.0\pm0.8$ and $n=4.2\pm1.2$.
}
\label{fig:L(n)_total}
\end{figure}

\begin{figure}
\epsscale{1.}
\plottwo{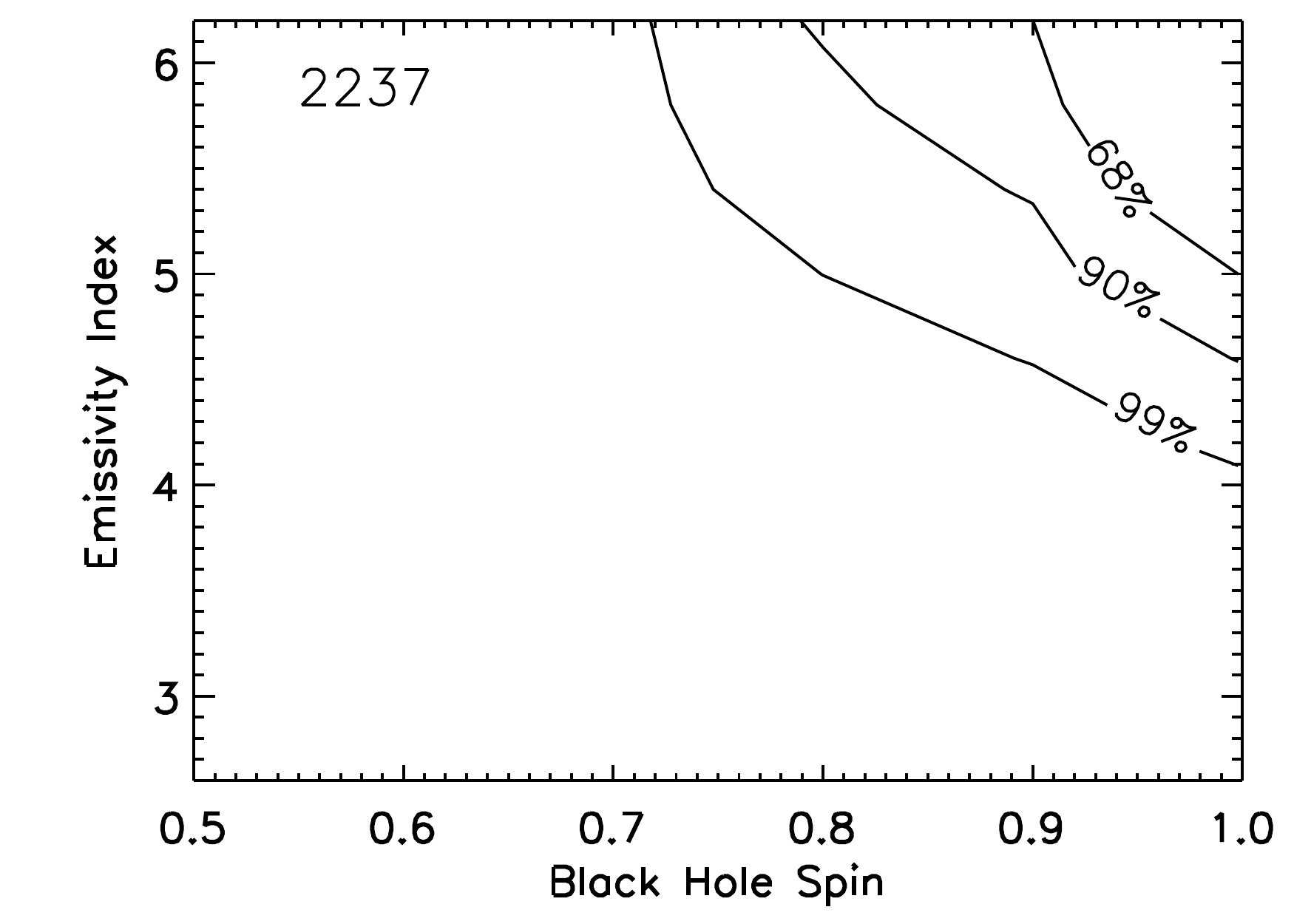}{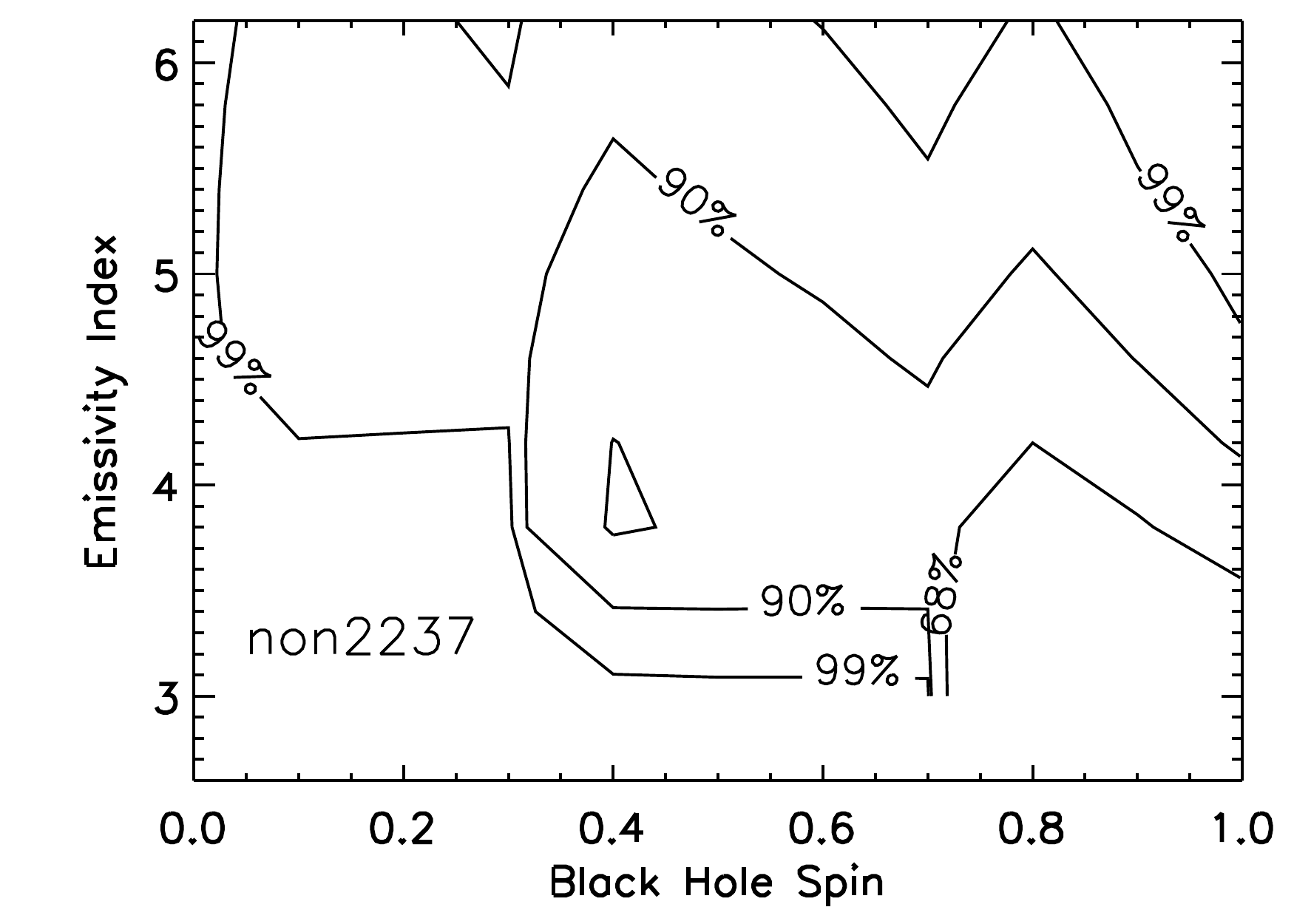}
\caption{68\%, 90\%, and 99\% confidence contours in the two-dimensional parameter space of $a$ and $n$ for \Q\ (left) and the joint sample of \QJ, \HEone, \SD, and \HEtwo\ (right).}
\label{fig:con}
\end{figure}

\begin{figure}
\epsscale{1.}
\plotone{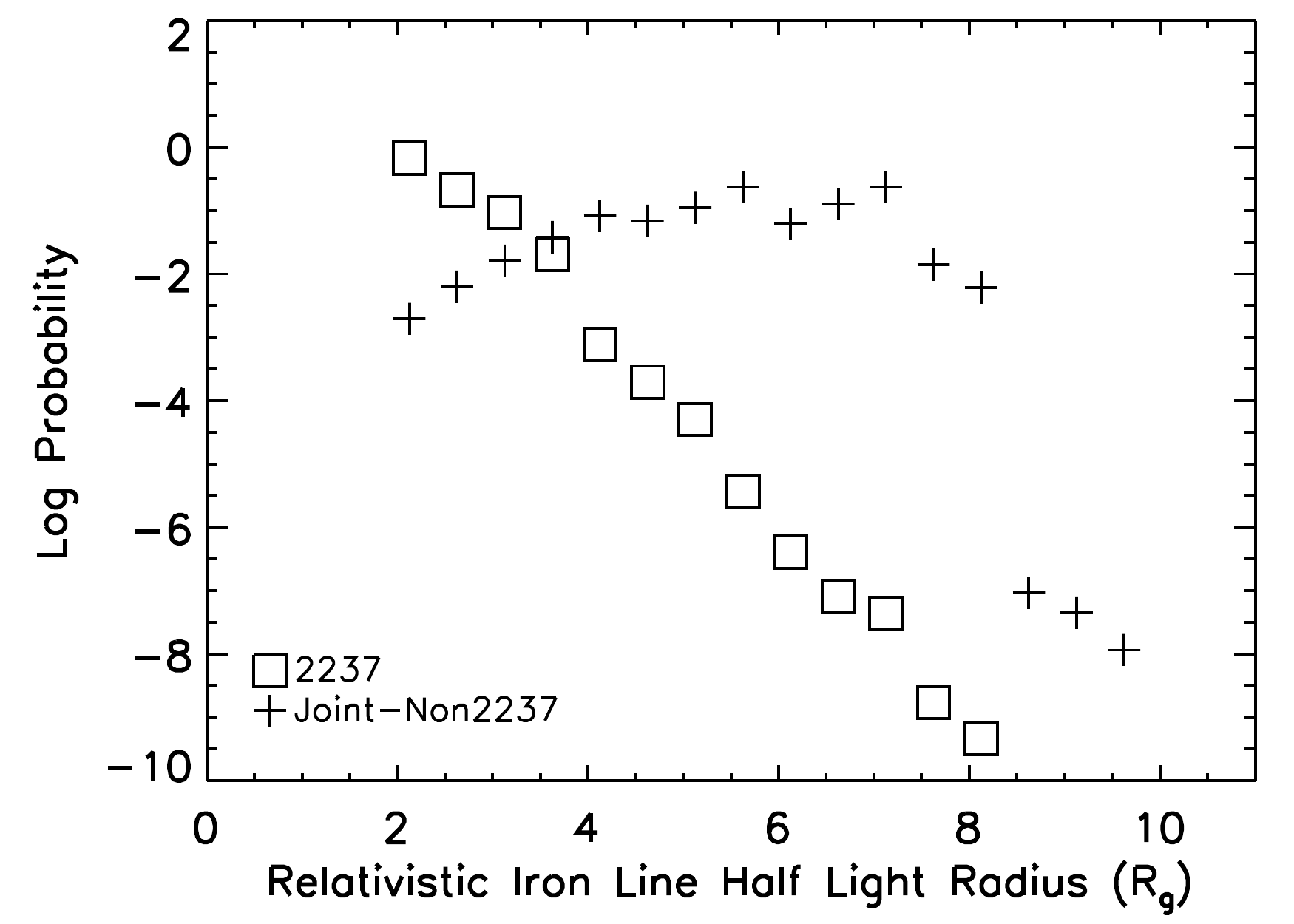}
\caption{Logarithm likelihood as a function \feka\ emission half light radius for \Q\ and the joint sample of \QJ, \HEone, \SD, and \HEtwo.  The 68\% confidence ranges of the half light radius are $< 2.4$ $r_g$ for \Q\ and 5.9--7.4 $r_g$ for the joint sample.}
\label{fig:hl}
\end{figure}

Figures~\ref{fig:L(a)_total} and \ref{fig:L(n)_total} show the marginalized probabilities for the spin and emissivity index parameters.
For \Q, we obtain tight constraints on both the spin and emissivity index parameters with well established probability peaks, and the 68\% and 90\% confidence limits for the spin parameter are $a > 0.92$ and $a>0.83$, respectively, where we linearly interpolate the probabilities to match the designated limits.
The corresponding 68\% and 90\% confidence limits for the emissivity index are $n > 5.4$ and $n > 4.9$ for \Q.
Compared to other targets, \Q\ has the longest exposure among the sample, the line EWs have small relative uncertainties, and the EW deviations from the Iwasawa-Taniguchi relation are large, and because of these factors, the constraints for \Q\ are strong. 
For the remaining four targets \QJ, \HEone, \SD, and \HEtwo, the individual constraints are weak. 
However, since the shapes of the probability distributions are similar (Figure~\ref{fig:L(a)_total} right),  we jointly constrain the remaining targets by multiplying the probability functions, yielding 68\% and 90\% limits of $a=0.8\pm0.16$ and $a > 0.41$ for the spin parameter and $n=4.0\pm0.8$ and $n=4.2\pm1.2$ for the emissivity index.
We need to remove \Q\ from the joint sample; otherwise the probabilities for the joint sample will be dominated by a single object.
We plot the two dimensional confidence contours of the two parameters for \Q\ and the remaining sample in Figure~\ref{fig:con}.
We can also bin the two-dimensional parameter space by the half-light radius after the Kerr lensing effect and calculate the corresponding probabilities in each bin.  Figure~\ref{fig:hl} shows the normalized probabilities as a function of \feka\ emission radius for \Q\ and the remaining joint sample in the logarithm scale. The half light radius are constrained to be $< 2.4$ $r_g$ and $<2.9$ $r_g$ (68\% and 90\% confidence) for \Q\ and in the range of 5.9--7.4 $r_g$ (68\% confidence) and 4.4--7.4 $r_g$ (90\% confidence) for the joint sample.

\section{DISCUSSION}
\label{sec:discussion}
Under the hypothesis that the higher average EW of lensed quasars for a monitoring sequence of observations is a microlensing effect, we explain the offset using a set of general relativistic corona, reflection, and microlensing models. We perform a microlensing analysis to obtain the likelihood as a function of the index of the emissivity profile for the reflection component and spins of black holes, which we have included a modest redshift evolution effect on the rest-frame EW of \feka\ lines, such that the spin values obtained are more conservative.

For the joint constraint from a sample of four targets, our analysis showed that the relativistic reflection region is more likely to have an emissivity index of $n=4.0\pm 0.8$ and a half light radius of 5.9--7.4 $r_g$ ($1 \sigma$), and therefore originates from a more compact region relative to the continuum emission region. 
This result confirms the previous qualitative microlensing argument that point towards the reflection region belonging to a more compact region \citep[e.g.,][]{chen2012}. 
The result also shows that the X-ray continuum cannot be a simple point source ``lamppost'' model, confirming the earlier analysis result of \citet{popovic2006}.
The spin value of the joint sample is constrained to be $a=0.8\pm0.16$.
This is in agreement with previous studies reporting high spin measurements \citep[e.g.,][]{reis2014, reynolds2014, mreynolds2014, capellupo2015, capellupo2017} either in the local or high redshift samples. 
For \Q, both the spin and emissivity index parameters are well constrained individually with \(a > 0.92\) and $n>5.4$ corresponding to 2.25--3 $r_g$ for spins between 0.9 and the maximal value.
Overall, our spin measurements favor the ``spin-up'' black hole growth model, where most of the accretion occurs in a coherent phase with modest anisotropies, especially for $z > 1$ quasars \citep[e.g.,][]{dotti2013, volonteri2013}. 

Since this paper uses the relative microlensing magnification as a signal to constrain the emissivity profiles of the reflection region, the technique only probes the \feka\ emission region comparable or smaller than the X-ray continuum emission regions. Emission lines originating from this compact region are theoretically predicted to have a broad line profile and with the peak energy varying with the inclination angle.  
The broad emission line widths, especially for \Q\ of $\sim$\,0.5\,keV and 4--8 $\sigma$ broad, and the range of line energies between 5.8--6.8\,keV, provide the confirmation of this theoretical expectation.
For the reflections that occur at much larger distances, at the outer portion of the accretion disk, disk wind, broad line region, or torus, they will result in a narrow \feka\ line that is not sensitive to this technique.  
It is also quite possible that our sample is biased because we selected our targets based on their strong microlensing signals at optical wavelengths.
However, since being microlensing active and having large spin values are independent, we do not see this bias will significantly affect our results.  
Furthermore, \HEtwo B was the only image with no detectable \feka\ features and was discarded in the microlensing analysis, suggesting that our somewhat limited exposure times were sufficiently large as to not introduce any non-detection bias.
This will not affect the microlensing constraints for \Q, and will only have limited effect on the joint sample results because it contributes less than a tenth of the sample.
\citet{reis2014} and \citet{mreynolds2014} fit a broad relativistic \feka\ line to the stacked spectra of gravitationally lensed quasars.  This technique assumes that the stacked \feka\ line profile resembles the un-lensed line profile; however, the \feka\ line peak is observed to be a variable between observations \citep{chartas2017}.  Although this technique has a different set of systematic uncertainties, the resulting constraints are quite similar to the analysis results from this paper.  For \Q, the line fitting method has yielded $a = 0.74^{+0.06}_{-0.03}$ \citep[90\% confidence,][]{mreynolds2014}, and the constraint in this paper is $a > 0.83$ (90\% confidence). Both studies show that \Q\ has a large spin values, while the analysis here points more to a maximal value.  The steep emissivity profiles measured in this paper are also broadly consistent with those measurements from local AGN, such as MCG-6-30-15 \citep{wilms2001,vf2004,miniutti2007},  1H0707$-$495 \citep{zoghbi2010, dauser2010}, and IRAS 13224$-$3809 \citep{ponti2010}.  These steep emission profiles can be resulted by combining the light bending, vertical Doppler boost, or ionization effects to produce slopes as steep as $n\sim7$ \citep{wilms2001,vf2004,fk2007,svoboda2012}.

Unfortunately, the spin measurement technique presented in this paper can only be used to analyze the small sample of targets whose X-ray spectra can be measured with sufficient signal-to-noise ratios using the current
generation of X-ray telescopes.
The next generation X-ray telescopes with an order of magnitude increase in the effective area will allow these measurements in a much larger sample.  Ideally, we need sub-arcsec angular resolutions to resolve the lensed images to increase the constraining power for the size and spin measurements.  However, a similar analysis can be applied to the total image of the lensed quasars, where the requirement for the angular resolution is less crucial, because the analysis relies on the time-averaged relative microlensing signals between the X-ray continuum and \feka\ emission regions.  In addition, quasar microlensing can induce variability in the polarization signals, especially the polarization angle \citep{chen2015b}, which can be detected by future X-ray polarization missions and put constraints on quasar black hole spins independently.

\acknowledgements 
We acknowledge the financial support from the NASA ADAP programs NNX15AF04G, NNX17AF26G, NSF grant AST-1413056, and SAO grants AR7-18007X, GO7-18102B. CWM is supported by NSF award AST-1614018.  We thank D.~Kazanas, L.\ C.~Popovic for helpful discussion and the anonymous referee for valuable comments.

\clearpage

\end{document}